\documentclass[a4paper,useAMS,usenatbib]{mnras}
\pdfoutput=1

\usepackage{graphicx}
\usepackage{setspace}
\usepackage{natbib}
\usepackage{color}
\usepackage{amsmath,amssymb}
\usepackage{times}
\usepackage{hyperref}[]

\title[Energy wrinkles and phase-space folds]{Energy wrinkles and phase-space folds of the last major merger}

\author[Belokurov et al]{Vasily  Belokurov$^{1,2}$\thanks{E-mail:vasily@ast.cam.ac.uk}, Eugene Vasiliev$^{1}$, Alis J. Deason$^{3,4}$, Sergey E. Koposov$^{5,1,6}$, \newauthor Azadeh Fattahi$^{3}$, Adam M. Dillamore$^{1}$, Elliot Y. Davies$^{1}$, and Robert J. J. Grand$^{7,8}$\\
$^1$Institute of Astronomy, Madingley Rd, Cambridge, CB3 0HA, UK\\ 
$^2$Center for Computational Astrophysics, Flatiron Institute, 162 5th Avenue, New York, NY 10010, USA\\
$^{3}$Institute for Computational Cosmology, Department of Physics, Durham University, South Road, Durham DH1 3LE, UK\\
$^{4}$Centre for Extragalactic Astronomy, Department of Physics, Durham University, South Road, Durham DH1 3LE, UK\\
$^{5}$Institute for Astronomy, University of Edinburgh, Royal Observatory, Blackford Hill, Edinburgh EH9 3HJ, UK\\
$^{6}$Kavli Institute for Cosmology, University of Cambridge, Madingley Road, Cambridge CB3 0HA, UK\\
$^7$Instituto de Astrof\'isica de Canarias, Calle Vía L\'actea s/n, E-38205 La Laguna, Tenerife, Spain\\
$^8$Departamento de Astrof\'isica, Universidad de La Laguna, Av. del Astrof\'isico Francisco S\'anchez s/n, E-38206, La Laguna, Tenerife, Spain
}

\begin{document}

%\date{August 2015}
%\pagerange{\pageref{firstpage}--\pageref{lastpage}} \pubyear{2015}

\maketitle

\label{firstpage}

\begin{abstract}
Relying on the dramatic increase in the number of stars with full 6D phase-space information provided by the {\it Gaia} Data Release 3, we discover unambiguous signatures of phase-mixing in the stellar halo around the Sun. We show that for the stars likely belonging to the last massive merger, the $(v_r,r)$ distribution contains a series of long and thin chevron-like overdensities. These phase-space sub-structures are predicted to emerge following the dissolution of a satellite, when its tidal debris is given time to wind up, thin out and fold. Additionally, the observed energy and angular momentum $(E, L_z)$ distribution appears more prograde at high energies, possibly revealing the original orbital angular momentum of the in-falling galaxy. The energy distribution of the debris is strongly asymmetric with a peak at low $E$ -- which, we surmise, may be evidence of the dwarf's rapid sinking -- and riddled with wrinkles and bumps. If these small-scale energy inhomogeneities have been seeded during or immediately after the interaction with the Milky Way, and are not due to the spatial restriction of our study, then making use of the $(v_r,r)$ chevrons to constrain the time of the merger becomes cumbersome. Nonetheless, we demonstrate that similar phase-space and $(E,L_z)$ sub-structures are present in numerical simulations of galaxy interactions, both in bespoke N-body runs and in cosmological hydrodynamical zoom-in suites. The remnant traces of the progenitor's disruption and the signatures of the on-going phase-mixing discovered here will not only help to constrain the properties of our Galaxy's most important interaction, but also can be used as a novel tool to map out the Milky Way's current gravitational potential and its perturbations.

\end{abstract}


\section{Introduction}

The most striking global feature of the Milky Way's stellar halo is the radial density break around $20-30$ kpc from the Galactic Centre \citep[see e.g.][]{Watkins2009,Deason2011,Sesar2011}. This dramatic steepening of the halo star counts was interpreted by \citet{Deason2013} to conclude that the accretion history of the Galaxy had been dominated by a single, ancient and massive merger. This hypothesis was tested with the arrival of the {\it Gaia} data \citep[][]{Gaia} which revealed the prevalence in the inner halo of the Milky Way of relatively metal-rich stars on highly eccentric orbits, attributed to a dwarf galaxy with a total mass of order of $10^{11} M_{\odot}$ accreted some 8-11 Gyr ago \citep[][]{Belokurov2018,Helmi2018}. This merger event, known today as {\it Gaia} Sausage/Enceladus (GS/E), has by now been mapped out in the chemical and kinematic space \citep[e.g.][]{Haywood2018,Mackereth2019,Necib2019,Lancaster2019,Das2020,Feuillet2021,Carrilo2022}. Taking advantage of the unprecedented quality of {\it Gaia}'s astrometry, the stellar halo break has been shown to be created by the apocentric pile-up of GS/E stars turning around on their orbits \citep[][]{Deason2018}. 

The global 3D shape of the Milky Way's inner halo has been charted with the RR Lyrae stars and shown to be triaxial with a major axis lying in the plane of the Galactic disk more or less aligned with the crossing of the orbital plane of the Magellanic Clouds \citep[][]{Iorio2018}. Subsequently, using the {\it Gaia} Data Release 2 proper motions, \citet{Iorio2019} demonstrated that the bulk of the RR Lyrae inside 30 kpc of the Galactic Centre move on similar, highly radial orbits and are likely all part of the same accretion event, namely the GS/E. Viewed projected onto the $x-z$ plane, the  intermediate axis of the inner stellar halo is slightly tilted out of the Galactic disk plane (see Figure 3 of \citealt{Iorio2019}). The Sun's position is only $\sim20^{\circ}$ off the intermediate axis, but this is enough to see the close side of the debris cloud spanning a larger projection on the sky compared to the one further away, in effect similar to the sky projection of the Galactic bar. The outer density contours show clear deviations from the simple ellipsoidal shape corresponding to the previously known debris ``clouds" \citep[][]{Simion2019,Balbinot2021}.

There have been several early signs that at least in the Solar neighbourhood, the GS/E tidal debris dominate the accreted portion of the stellar halo \citep[e.g.][]{Brook2003,Meza2005,Nissen2010}. Today, the dominance of the GS/E debris locally and throughout the inner MW halo has been well established. For example, most recently, \citet{Myeong2022} demonstrated that the GS/E is the only significant accreted component amongst local stars on eccentric orbits. Curiously, three other halo components they find are all of in-situ nature, i.e. born in the MW proper. These include {\it Splash}, i.e. heated high-$\alpha$ disk of the Galaxy \citep[][]{Bonaca2017,Gallart2019,Dimatteo2019,Belokurov2020}, {\it Aurora}, i.e. the pre-disk quasi-spheroidal early Galaxy \citep[][]{Aurora,Conroy2022}, as well as {\it Eos}, a new in-situ halo component linked to the low-$\alpha$ disk \citep[see][]{Myeong2022}.

The census of the most significant substructures in the Solar neighbourhood reported by \citet{Myeong2022} can be interpreted with the detailed analysis of a massive satellite in-fall. \citet{Amorisco2017} and \citet{Vasiliev2022} demonstrate that if the satellite-host mass ratio is sufficiently high, for a certain range of central densities of the two galaxies, the interaction proceeds in an unexpected and previously poorly understood way. Instead of sinking in the host's potential with ever-increasing circularity as predicted by the Chandrasekhar's Dynamical Friction (DF) prescription \citep[][]{Chandra1943,Jorge2004}, the satellite's orbital eccentricity rapidly ramps up causing it to stall, drop to the centre of the host and fall apart in an accelerated, explosive fashion. \citet{Vasiliev2022} show that the satellite's orbital radialization is caused by a complex mix of distortion and reflex motion of the host as well as satellite's self-friction. These factors, not considered in the classical DF picture, help the satellite to sink and disrupt faster, thus inundating the Solar neighbourhood with its debris. The satellite's and the host's properties need to be just right for the radialization to happen efficiently. If, for example, the central density is too low, the disruption happens too quickly, before the satellite arrives to the heart of the Milky Way. Dialling the densities up will reduce the satellite's mass loss and therefore will lessen the self-friction. It will also subdue the reflex motion and consequently will slow down the radialization or even completely reverse it. Profound radialization can thus be considered a giveaway that the satellite survived more or less intact (at least its stellar component) until arriving close to the Solar neighbourhood. 

For a progenitor whose orbit does not evolve significantly during the disruption, the tidal debris behaviour is best summarised in the space of integrals of motion \citep[][]{Johnston1998, HelmiWhite1999}. Viewed either in the action space \citep[][]{Eyre2011} or in the energy and angular momentum space, stars stripped from the in-falling galaxy, form characteristic bow-tie shapes \citep[see e.g. Figure~1 in][]{Gibbons2014}. The opposite ends of the tie are the leading and the trailing debris, with correspondingly lower and higher energy compared to the progenitor. While these shapes get somewhat smeared by both the host's evolution over time and the measurement errors, they remain recognisable and can be used to decipher the accretion history of the Galaxy \citep[see][]{Helmi2000}. The situation is much more complex for a rapidly sinking satellite. In this case, how fast the satellite plunges in the host's potential and the rate at which it loses stellar mass and angular momentum will now control the appearance of its tidal debris. While for a stationary orbit, the debris unbound at each stripping episode contributes to the same footprint in e.g. $(E,L_z)$ space, the bow-tie shape of a plunging satellite is constantly distorted and dragged to lower energies. As a result, this leads to the formation of a twisted and stretched (along $E$) column of debris \citep[see e.g.][]{Koppelman2020,Amarante2022,Khoperskov2022}. It is therefore a generic expectation that the energy distribution of the tidal debris of a satellite on a rapidly decaying orbit should contain multiple bumps and wrinkles corresponding to the individual stripping episodes along the satellite's journey down the potential well. Moreover, each such episode should produce at least two energy pile-ups, corresponding to the leading and trailing debris.

\begin{figure*}
  \centering
  \includegraphics[width=0.99\textwidth]{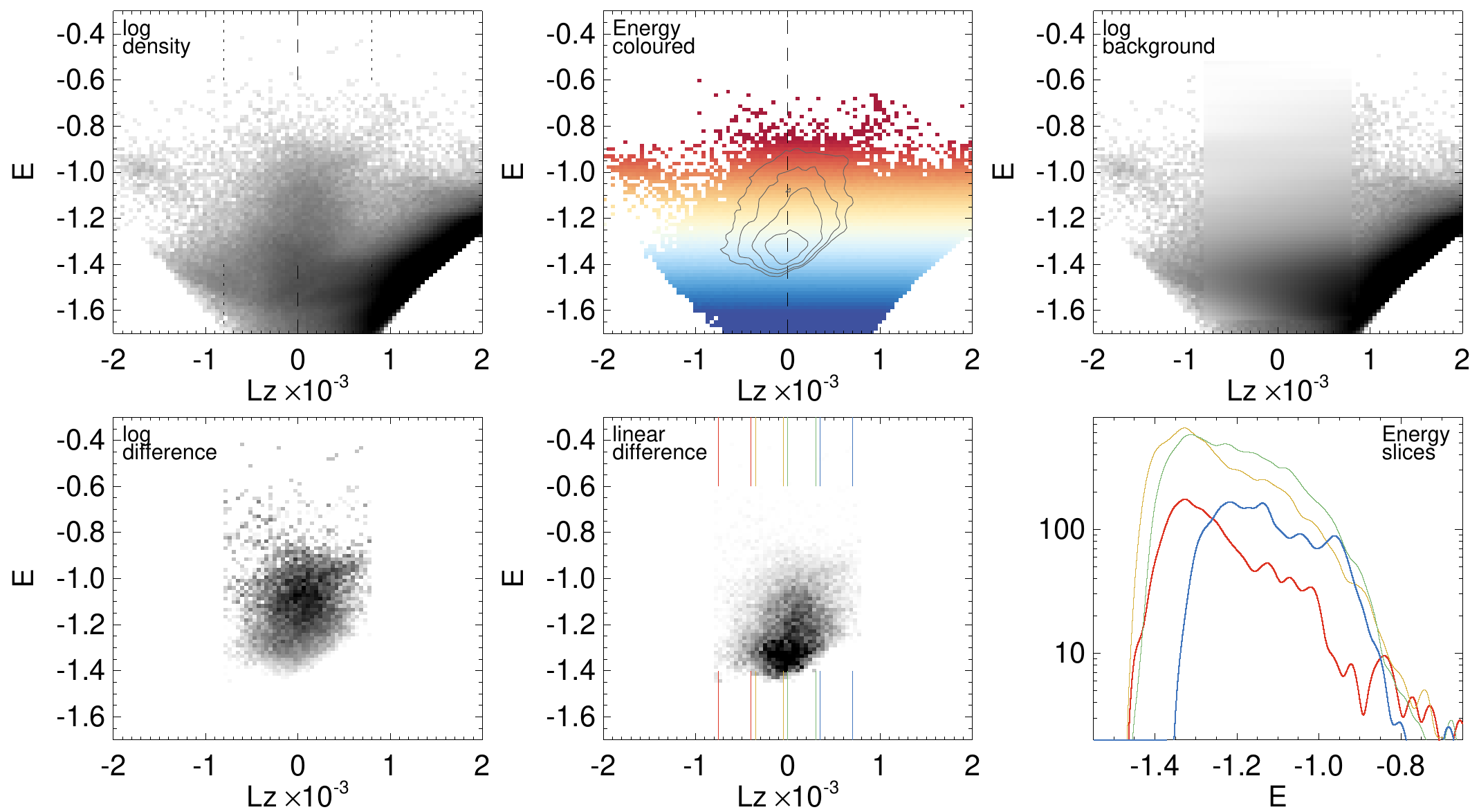}
  \caption[]{$E$-$L_z$ properties of the local {\it Gaia} DR3 RVS stars. {\it Top, left:} Logarithm of stellar density in the space of total energy $E$ and the angular momentum's vertical component $L_z$. Energy is in units of $10^5\, \mathrm{km}^2\,\mathrm{s}^{-2}$. The $E, L_z$ region shown is binned into $100\times100$ equal size pixels. Dotted vertical lines mark the region for which the background density is estimated (see the right panel in this row). {\it Top, middle:} Same as previous panel colour-coded according to the median energy. {\it Top, right:} Logarithm of the background density estimate. {\it Bottom, left:} Same as the top left but with the background estimate subtracted. The bulk of the remaining over-density is the GS/E debris cloud. {\it Bottom, middle:} Linear density residual (also shown as density contours above, in the top middle panel). Note the avocado-shaped $(E,L_z)$ distribution of the GS/E debris. If not caused by the {\it Gaia} DR3 RVS selection function or issues with the background subtraction, the bottom-heavy distribution of energy may be an observational evidence for the progenitor's rapid sinking in the host's potential. Coloured vertical lines mark four $L_z$ bins for which GS/E energy profiles are shown in the right panel. {\it Bottom, right:} Slices though the GS/E debris (as represented by the residual density shown in the middle panel) at fixed $L_z$.
  }
   \label{fig:elz}
\end{figure*}
Once packs of stripped stars are deposited into the host's gravitational potential, in principle, their distribution in the space of integrals of motion barely changes, but their phase-space density constantly evolves. Small differences in stars' orbital frequencies eventually translate into orbital phase offsets that accumulate with time in the process known as phase-mixing. As the stellar debris cloud spreads over the Milky Way, the increase of its spatial extent is balanced by the thinning out of the velocity distribution, keeping the phase-space density constant in accordance with Liouville's theorem \citep[see][]{HelmiWhite1999}. As the debris cloud continues to stretch in the phase-space, it eventually folds onto itself leading to a formation of a winding pattern, which resembles a spiral for orbits close to circular. Such phase-space spiral was uncovered recently in the disk stars around the Sun and is believed to be produced by a relatively recent perturbation of the Galactic disk by a massive body \citep[see][]{Antoja2018}. For an eccentric orbit, as viewed for example in the phase-space spanned by the Galactocentric spherical polars $v_r$ and $r$, the folds of the debris cloud appear as nested chevrons, but topologically are nonetheless a spiral, albeit severely distorted \citep[e.g. Figure~6 in][]{Quinn1984}. In the early stages of  phase-mixing, the chevrons expand radially, while creating sharp, caustic-like density enhancements around their apocentric radii known as shells \citep[][]{Sanderson2013}. Each stripping episode creates its own set of connected folds. As folds seeded at different times stretch, thin out and expand, they tend to run into each other and overlap \citep[see][]{DongPaez2022}, creating something akin to super-chevrons.

Thanks to phase-mixing, the coarse-grained phase-space substructure, i.e. lumps left behind by each episode of tidal stripping, is broken down into finer folded filaments and is, eventually, erased. At later stages, the phase-space density appears more uniform but is in fact made up of tightly packed series of folds. These may be difficult or impossible to tell apart when the entirety of the tidal debris is analysed. However, if a small spatial region, e.g. the Solar neighbourhood, is isolated, individual folds can be revealed \citep[see][]{McMillan2008,Gomez2010}. This is because given the same (or similar) time of release from the parent, only stars with certain orbital frequencies will have completed the right number of orbits around the Milky Way to enter the region around the Sun. Therefore, spatially localized views of the tidal debris in either phase-space or any space spanned by integrals of motion is always expected to be lumpy. The size of the clumps and the spacing between them can in fact be used to deduce the time of accretion \citep[see][]{McMillan2008,Gomez2010}. Alternatively, the gravitational potential can be constrained by the requirement to bring the integrals of motion clumps into a sharp focus \citep[][]{DongPaez2022}.

In this paper, we report the first observational evidence for a sequence of folds in the phase-space of the Milky Way's stellar halo. After introducing our input dataset in Section~\ref{sec:data}, we explore two complementary views on the phase-space distribution of halo stars: energy vs. the $z$-component of angular momentum (Section~\ref{sec:energy}) and Galactocentric distance vs. radial velocity (Section~\ref{sec:phase-space}). We then examine the dependence of the phase-space sub-structure on the stellar metallicity to reveal the presence of the in-situ halo signal. We compare the {\it Gaia} DR3 observations to numerical simulations in Section~\ref{sec:sims}, and summarize our findings and their interpretation in Section~\ref{sec:conc}. 

\section{Data and sample selection}
\label{sec:data}

\begin{figure*}
  \centering
  \includegraphics[width=0.99\textwidth]{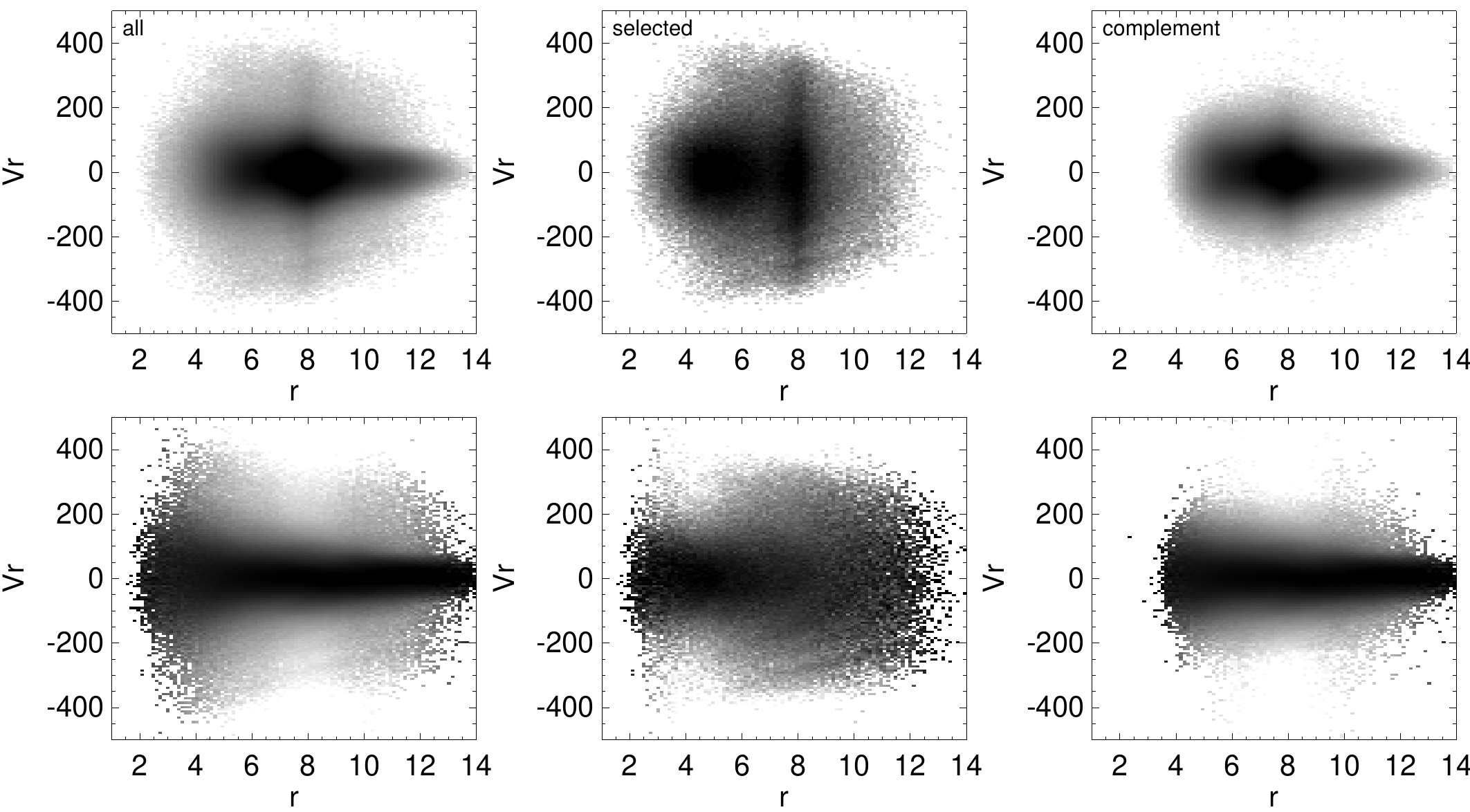}
  \caption[]{Solar neighbourhood view of the phase-space spanned by the spherical polar coordinates $v_r$ and $r$. There are 100 (150) pixels along $r$ ($v_r$) dimensions. {\it Top row:} Logarithm of the density of stars. {\it Bottom row:} Logarithm of the column-normalised density of stars. {\it Left:} All stars. {\it Middle:} Stars with $|L_z|<0.5\times10^3$. {\it Right:} Stars with $L_z>10^3$. Note the clear striation pattern appearing in the middle column.}
   \label{fig:phsp1}
\end{figure*}

We use data from the Radial Velocity Spectrograph (RVS) sample \citep[][]{gdr3_rvs} of the Data Release 3 from the {\it Gaia} space observatory \citep[][]{Gaia}. The astrometric solutions for these stars were provided previously \citep[][]{Lindegren2021} as part of the {\it Gaia} Early DR3 \citep[][]{gaia_edr3}. We use geometric distances as estimated by \citet{BJ2021}, however switching instead to the inverse of the parallax does not change our results noticeably. We apply only basic selection cuts, i.e we require that stars have a  relative precision better than 10\% ( $\varpi/\sigma_{\varpi}>10$), and that they lie not too far from the Sun at $D<15$ kpc, leaving $\sim$25.9 million objects out of the original $\sim$33.7 million with non-zero line-of-sight velocities. Additionally, we remove stars projected within 1.5 degree distance from the centres of known globular clusters within 5 kpc of the Sun. Our final sample size is $\sim$25 million stars with full 6-D phase-space information. Converting the observed heliocentric stellar coordinates into the Galactocentric left-handed reference frame, we assume that the Sun is at $X=R_{\odot}=8$ kpc from the Galactic Centre \citep[although a slightly larger value was recently reported by][]{GRAVITY2022} and lies in the Galactic plane, at $Z_{\odot}=0$. Following \citet{Drimmel2022}, we assume that the Sun's velocity is $v_{\odot}=\{-9.3, 251.5, 8.59\}$ km s$^{-1}$. We have checked that changing any of the Galactic parameters within their systematic uncertainties does not change any of the conclusions reported below.

\begin{figure*}
  \centering
  \includegraphics[width=0.99\textwidth]{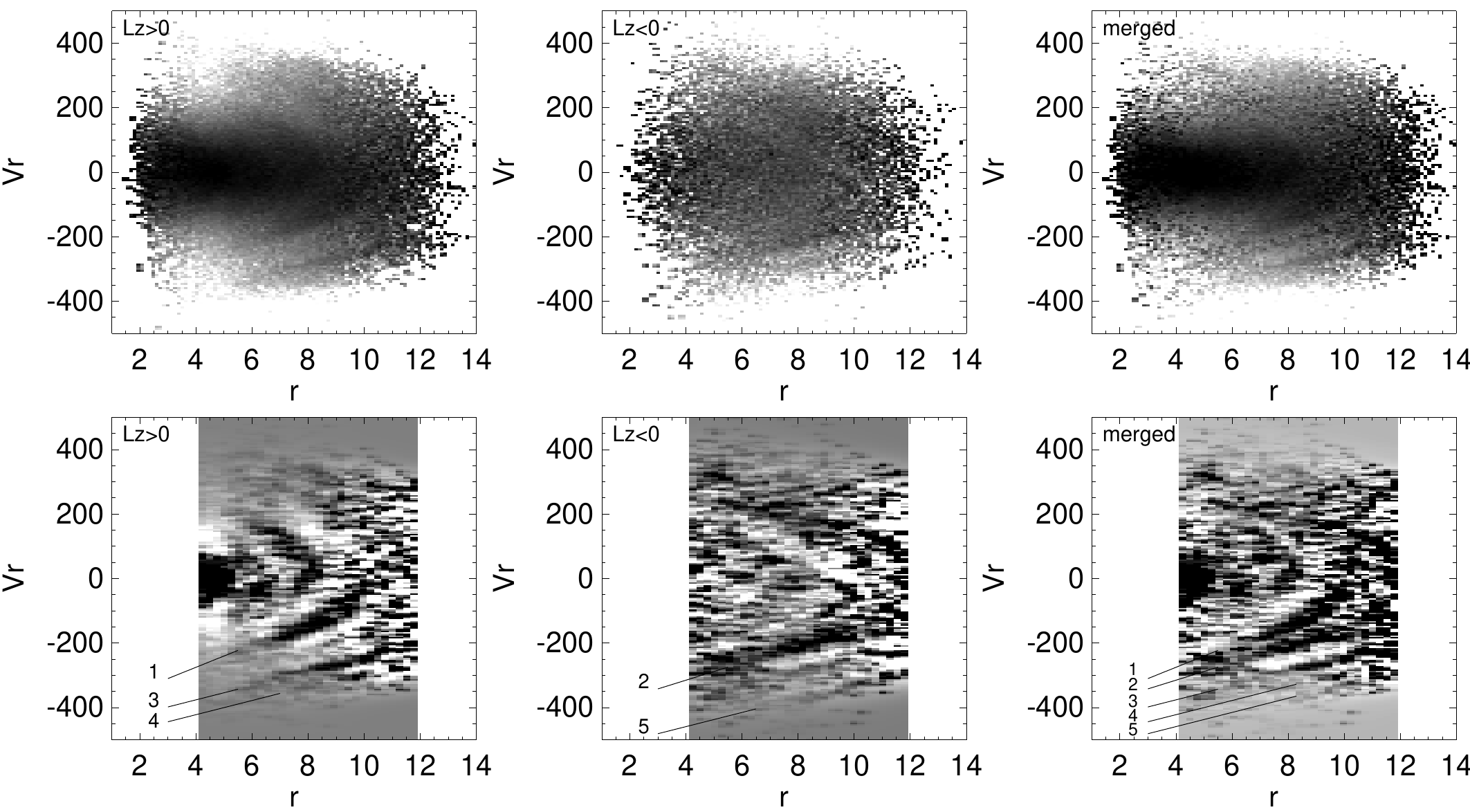}
  \caption[]{GS/E phase-space folds in the Solar neighbourhood (for stars within $\approx5$ kpc from the Sun). {\it Top row:} Logarithm of column-normalised density in $v_r,r$ space. There are 100 (150) pixels along $r$ ($v_r$) dimensions. {\it Bottom row:} As above but with the smooth background subtracted to reveal the GS/E folds. There 50$\times$200 pixels. The background is the density distribution convolved with a Gaussian kernel with a FWHM of 9 and 15 pixels for positive and negative $L_z$ selections, respectively. {\it Left:} Stars with $L_z>0$. {\it Middle:} Stars with $L_z<0$. {\it Right:} Combination of the previous two panels. Note that the striation pattern changes depending on the angular momentum selection. Chevrons 1 to 5 are marked. Note that the top right panel while being similar is not identical to the bottom centre panel of Figure~\ref{fig:phsp1}.}
   \label{fig:phsp2}
\end{figure*}
\begin{figure*}
  \centering
  \includegraphics[width=0.99\textwidth]{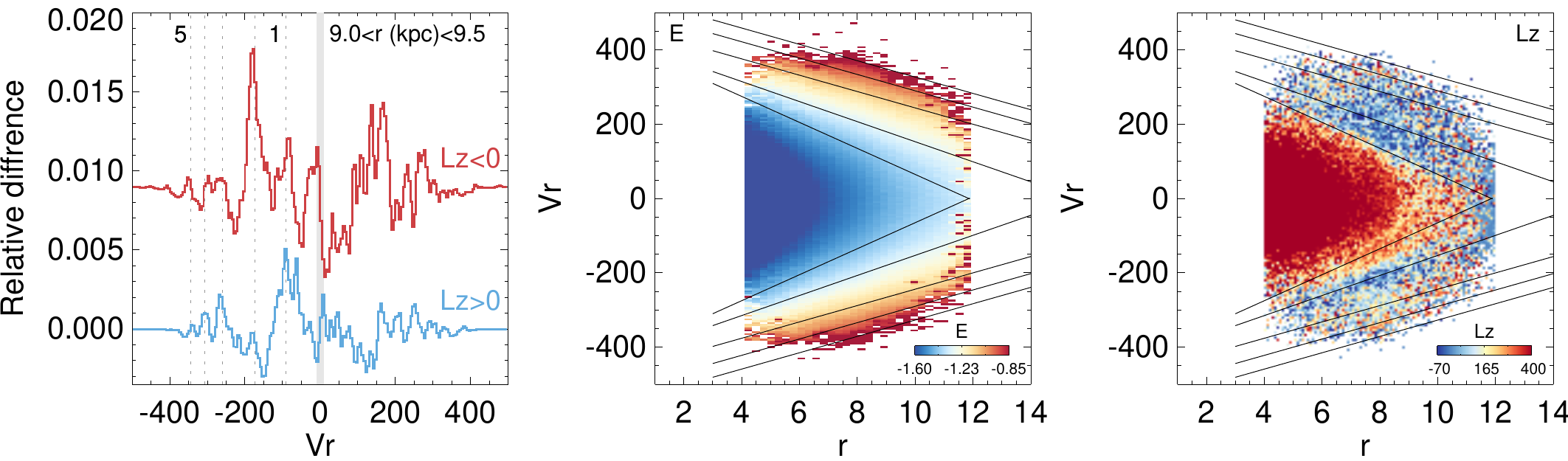}
  \caption[]{Properties of the local GS/E phase-space folds. {\it Left:} Slices through the background-subtracted density distribution shown in Figure~\ref{fig:phsp2} at $9<r$(kpc)$<9.5$. Red (blue) histograms are for stars on halo-like orbits with $L_z<0$ ($L_z>0$). Vertical dotted lines show predicted locations of the 5 chevrons using linear $(v_r,r)$ approximation. All 5 chevrons are detected, albeit with different strength, in both $L_z<0$ and $L_>z$ samples. {\it Middle:} $v_r, r$ space coloured by the median energy, using the same colour scheme as in the top centre panel of Figure~\ref{fig:elz}). {\it Right:} $v_r, r$ space coloured by the median of the vertical component of the angular momentum, $L_z$. Note the alternating pattern of positive and negative $L_z$.}
   \label{fig:phsp3}
\end{figure*}

\section{Energy wrinkles}
\label{sec:energy}
 Figure~\ref{fig:elz} presents the behaviour of the local {\it Gaia} DR3 RVS stars in the space spanned by the total energy $E$ and the vertical component of the angular momentum $L_z$. Total stellar energies $E$ are computed using a three-component (bulge, disk and DM halo) Galaxy potential similar to the \texttt{MWPotential2014} of \citet{Bovy2015} but with the DM halo's mass of $M_{\rm vir}=10^{12} M_{\odot}$  instead of $0.8\times 10^{12} M_{\odot}$. We use a slightly higher concentration of the DM halo ($c=19.5$) to match the circular velocity at the Solar radius $v_{\rm circ}=235$ km s$^{-1}$. Energy is in units of $10^5\,\mathrm{km}^2\,\mathrm{s}^{-2}$. The top left panel of the Figure shows the logarithm of the stellar density distribution with a prominent vertical structure around $L_z=0$ corresponding to the GS/E tidal debris. A noticeable retrograde clump at $L_z<-10^3$ and $E>-1$ is the Sequoia structure \citep[][]{Myeong2019,Matsuno2019}. Note that adopting a different gravitational potential for the Galaxy may shift the patterns discussed below up and down slightly but is not going to change our overall conclusions.

To reveal the density variations across the GS/E debris cloud we estimate the behaviour of the smooth (well mixed) background as follows. In the $L_z$ range marked by two vertical dotted lines ($|L_z|<0.85\times10^3$, left panel of Figure~\ref{fig:elz}) the density is linearly interpolated using the values just outside the region of interest. To smooth the background variation, for each $E$ bin considered, we model the average of 9 $L_z$ profiles, i.e. additionally considering 4 rows of pixels above and below the current one. The background represents the well-mixed component of the local stellar halo. The bulk of this background component is contributed by the in-situ halo for which the assumption of mixedness may well be a good one: the in-situ halo stars are either already more phase-mixed (e.g. {\it Splash}) or has had the longest to phase-mix (e.g. {\it Aurora}). Subtracting the distribution shown in the right panel of the top row of Figure~\ref{fig:elz} from that shown in the left gives the difference of the logarithms of stellar densities (data-background) displayed in the bottom left panel. The bottom middle panel of the Figure shows the linear over-density residual. Note that given the choice of the size of the interpolated region, the background estimate is only valid down to energies of order of $E\approx-1.65$. Curiously, the residual corresponding to the GS/E only reaches $E\approx-1.4$. This could indicate a genuine drop in the number of GS/E stars at low energies, or, alternatively, problems with our background estimate below $E=-1.4$. The top right panel of Figure~\ref{fig:elz} indicates that the latter is indeed possible as the background quickly gets quite complex at low energies due to i) the contribution of the in-situ halo and ii) the {\it Gaia} DR3 RVS selection biases. If there are any stars belonging to the merger at lower energies they are not contributing to the GS/E signal analysed in this Section. The majority of other known (lower mass) halo sub-structures do not contribute many stars to the $L_z$ range of interest \citep[e.g.][]{Myeong2018,Myeong2019,Koppelman2019b}.

In the $E, L_z$ space, the 2-D GS/E over-density (Figure~\ref{fig:elz}, bottom row, left and middle; also contours in the top middle panel) has an elongated and inclined shape: its top portion (at high energy) leans towards $L_z>0$, while its bottom portion is more or less symmetric with respect to $L_z=0$, with a slight preference towards $L_z<0$. We interpret the inclination of the debris cloud in Section~\ref{sec:mergersim} and link it to the original angular momentum of the satellite at the in-fall. The debris cloud is also lumpy with a number of small-scale features such as i) a core at $L_z=0$ and $E\approx-1.4$, i.e. the bottom edge of the energy distribution, and ii) an energy depletion around $E\sim-1.1$, i.e. approximately through the middle of the over-density; this dent is most visible in the prograde portion of the cloud, i.e. at $L_z>0$. The shape of the $(E,L_z)$ overdensity resembles an avocado, with the bulk of the stars residing in the ``pit" in the broad bottom part of the debris cloud. The asymmetric, lopsided distribution of GS/E energies could potentially be an artefact of the background subtraction where some of the sharp density variations (to do with the contribution of the in-situ halo and the {\it Gaia} DR3 RVS selection function) have not been absorbed by the model. However, on close inspection, in the $(E, L_z)$ distribution shown in the top left panel of the Figure, no clear overdensity centered on $L_z=0$ below $E\approx-1.4$ is visible. Therefore the bottom-heavy energy distribution may be a genuine feature of the GS/E debris cloud and a direct observational evidence for progenitor's rapid sinking in the host's potential (see Section~\ref{sec:mergersim} for a comparison with a tailored N-body simulation).

The bottom right panel of Figure~\ref{fig:elz} shows four energy histograms (smoothed by convolving with an Epanechnikov kernel with a size of 1.3 pixel) corresponding to four slices through the GS/E cloud along $E$ in the $L_z$ bins marked with coloured vertical lines in the previous panel. All four histograms are noticeably asymmetric with peaks at lower energy levels, supporting the avocado picture introduced above. Moreover, the peak of the most prograde slice (blue) is at significantly higher energy compared to the most retrograde slice (red), confirming the tilted shape of the debris cloud in the  $(E, L_z)$ space. While the two profiles corresponding to the lowest $|L_z|$ are relatively smooth (yellow and green), the other two distributions corresponding to the most retrograde (red) and the most prograde (blue) edges of the GS/E debris show a number of wiggles. For example, the retrograde slice peaks just below $E\sim-1.3$ but has noticeable wrinkles (changes of curvature) between $E\sim-1.25$ and $E\sim-1$, together with an additional bump around $E\approx-0.8$. The prograde edge is clearly bimodal with bumps at $E\sim-1.15$ and $E\sim=-0.95$, on either side of the energy dent mentioned above and several additional wrinkles.

There are multiple reasons why the energy distribution can be wrinkled. Major overdensities are seeded at different energy levels as the disrupting satellite sinks in the host potential, producing at least two distinct energy clumps at each stripping episode. Additionally, if the tidal debris has had time to phase-mix, limiting stars to a relatively small spatial region would carve out portions of the energy distribution. Finally, subsequent rapid and strong perturbations of the Galaxy's potential can add extra wrinkles to the already complicated picture. We discuss some of these phenomena in Section~\ref{sec:sims} below.

\section{Phase-space folds}
\label{sec:phase-space}

\begin{figure}
  \centering
  \includegraphics[width=0.49\textwidth]{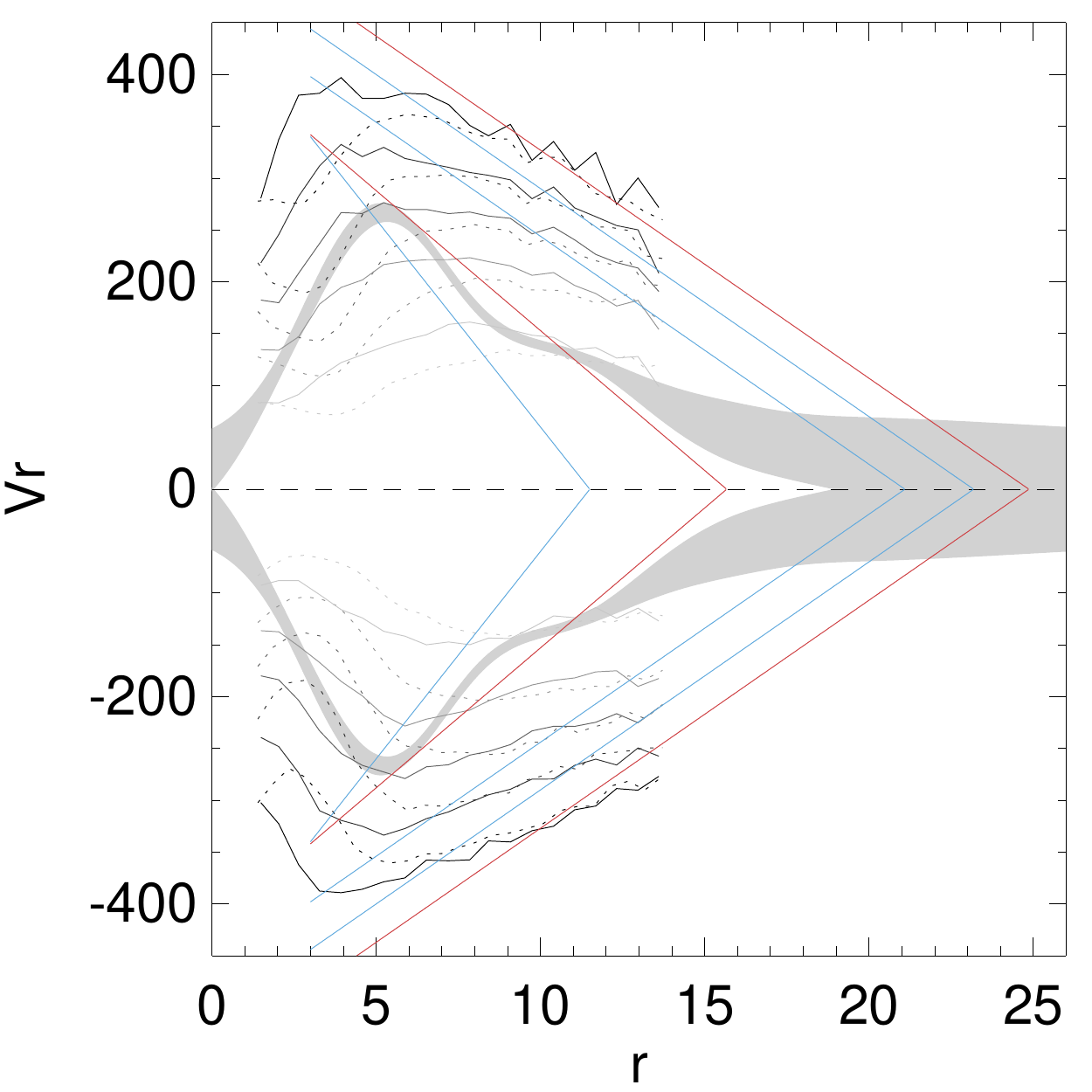}
  \caption[]{Phase-space structure of the GS/E debris. Red and blue lines mark the locations of the phase-space folds detected in $L_z>0$ and $L_z<0$ GS/E samples. Thin lines show the 0.5,2,5,10, 20 and 99.5, 98, 95, 90, 80 percentiles of the $V_r(r)$ distribution for stars with $|L_z|<0.5\times10^3$. The mean $V_r$ error is subtracted in quadrature. The grey band gives the comparison with the model of \citet{Iorio2021}}
   \label{fig:vrr_max}
\end{figure}

We now turn to the analysis of the phase-space, which does not depend on the assumptions about the potential. %\ev{probably need to discuss the relation between this and the previous spaces in more detail somewhere}. 

Figure~\ref{fig:phsp1} shows the $(v_r, r)$ phase-space behaviour of the high-quality sample of the {\it Gaia} DR3 RVS stars in the Solar neighbourhood. Here the top (bottom) row gives the logarithm of stellar density (column-normalised stellar density). The left column shows the phase-space density of all stars in the sample, while the middle panel corresponds to the stars selected to have $|L_z|<0.7\times10^{3}$, matching the properties of the GS/E debris as discussed in the previous section. This angular momentum selection gives $\approx$270,000 stars on eccentric (halo-like) orbits. The right panel shows the comparison sample with $|L_z|>10^{3}$. Hints of a striation pattern are already visible in the left column where low-level overdensities and underdensities move diagonally from high $|v_r|$ at low $r$ to low $|v_r|$ at high $r$. This pattern is amplified in the middle column where the stars within the GS/E's range of $L_z$ are selected. We interpret these phase-space feature as folds of the GS/E tidal debris as it stretches and winds up due to phase-mixing in the MW gravitational potential. The patterns in the first two columns can be compared to the right column where the stars at higher $L_z$ are shown. No obvious striation is visible for stars in the comparison sample in the right, indicating that the over- and under-densities are not an artefact of {\it Gaia} DR3 (e.g. due to the RVS selection function).

Figure~\ref{fig:phsp2} splits the GS/E sample into two portions, one with $L_z>0$ (left column) and one with $L_z<0$ (middle column). The logic of splitting the signal into two samples with different angular momentum is two-fold. First, the leading and trailing debris typically have distinct enough $E,L_z$ properties. Second, and most importantly, the in-situ halo contamination is a strong function of $L_z$, decreasing below $L_z=0$ (see the top right panel of Figure~\ref{fig:elz}). The difference in the amount of in-situ halo stars in $L_z>0$ and $L_z<0$ samples is obvious in the first two panels of the top row of Figure~\ref{fig:phsp2}. The left panel shows a prominent overdensity with $|v_r|<150$ km s$^{-1}$ corresponding to the {\it Splash} (with some contribution from {\it Aurora}), which is almost invisible in the middle panel ($L_z<0$).

The bottom row of Figure~\ref{fig:phsp2} shows the phase-space density after subtraction of a smooth background, similar to the view of the disk's phase-space overdensity pattern presented in e.g Figure 6 of \citet{Laporte2019}. The GS/E phase-space folds, i.e. quasi-linear over-dense and under-dense chevron-like regions are clearly visible in the bottom row of Figure~\ref{fig:phsp2}.  In the $L_z>0$ view (bottom left), there are two families of chevrons: those limited to $V_r<150$ km s$^{-1}$ and $r<8$ kpc and those with higher radial velocity amplitude present across the entire range of $r$. Because the first family is not detected in the $L_z<0$ sample we attribute these folds to the in-situ stellar halo (see Section~\ref{sec:sims} for further discussion) and focus the discussion in this Section on the high amplitude chevrons that we number 1 to 5 in the bottom row of Figure~\ref{fig:phsp2}. It is clear that while most of these high amplitude chevrons are present in both the $L_z>0$ and the $L_z<0$ views, their relative strength is different in the two samples. Chevrons 1, 3 and 4 are best seen in the $L_z>0$ view (left column), while chevrons 2 and 5 stand out more clearly above the background in the $L_z<0$ picture (middle panel). Curiously, in terms of the clarity of signal, the negative $v_r$ (moving towards the Galactic Centre) portions of the chevron pattern appear more coherent compared to their positive $v_r$ counterparts in both $L_z>0$ and $L_z<0$ samples. The right panel of the bottom row of the Figure combines the $L_z>0$ and the $L_z<0$ views and shows at least five clear and tightly packed phase-space folds.

Figure~\ref{fig:phsp3} summarises the properties of the GS/E phase-space substructure. The left panel of the Figure gives a slice (at $9.0<r$(kpc)$<9.5$) through the background-subtracted density distributions shown in the bottom row of Figure~\ref{fig:phsp2}. The corresponding velocities of each fold (as approximated by straight lines in the bottom row of Figure~\ref{fig:phsp2}) are shown as vertical dotted lines. At negative $v_r$, there is a perfect correspondence both between $L_z>0$ and $L_z<0$ slices and the predicted locations of the folds (vertical dotted lines). As mentioned above, all five chevrons are detected in both slices. Moreover, chevron 1 is bifurcated into two in the $L_z>0$ view, with only one portion of the bifurcation present in the $L_z<0$ slice, which could be due to the presence of the in-situ stars in the $L_z>0$ sample. The middle panel of Figure~\ref{fig:phsp3} shows the  $(v_r, r)$  phase-space colour-coded by the median energy of the stars in each pixel. As expected chevrons run through the phase-space at approximately constant energy. The strongest chevrons 1 and 2 have energies around $-1.4$ and $-1.2$ roughly matching the locations of the strongest peaks in the $E$ distributions shown in the bottom right panel of Figure~\ref{fig:elz}.

Finally, the right panel shows the same space colour-coded by the median $L_z$ and reveals the alternating positive/negative $L_z$ pattern of the GS/E folds detected. As the middle panel demonstrates, the phase-space follow (approximately) the lines of constant energy \citep[see][for an extensive discussion]{DongPaez2022}. Using a very crude straight line fit through the detected folds as shown in the bottom row of Figure~\ref{fig:phsp2}, we can estimate the apocentric distances of the stars that make up the chevrons. Starting from the lowest energy chevron, the stars in the five folds reach their apocentres around 11.5, 15.5, 21, 23 and 25 kpc from the Galactic centre. Pile-ups of the tidal material in the packs of folds can cause noticeable changes in the stellar halo radial density profile. The last three folds appear to be associated with the stellar halo break detected previously \citep[see e.g.][]{Watkins2009,Deason2011,Sesar2011}. The chevrons 1 and 2 are likely to introduce another break in the stellar halo density profile, at around $10<r$(kpc)$<15$. This is consistent with the predictions of a "double-break" halo in \citet{Naidu2021}.

\begin{figure}
  \centering
  \includegraphics[width=0.49\textwidth]{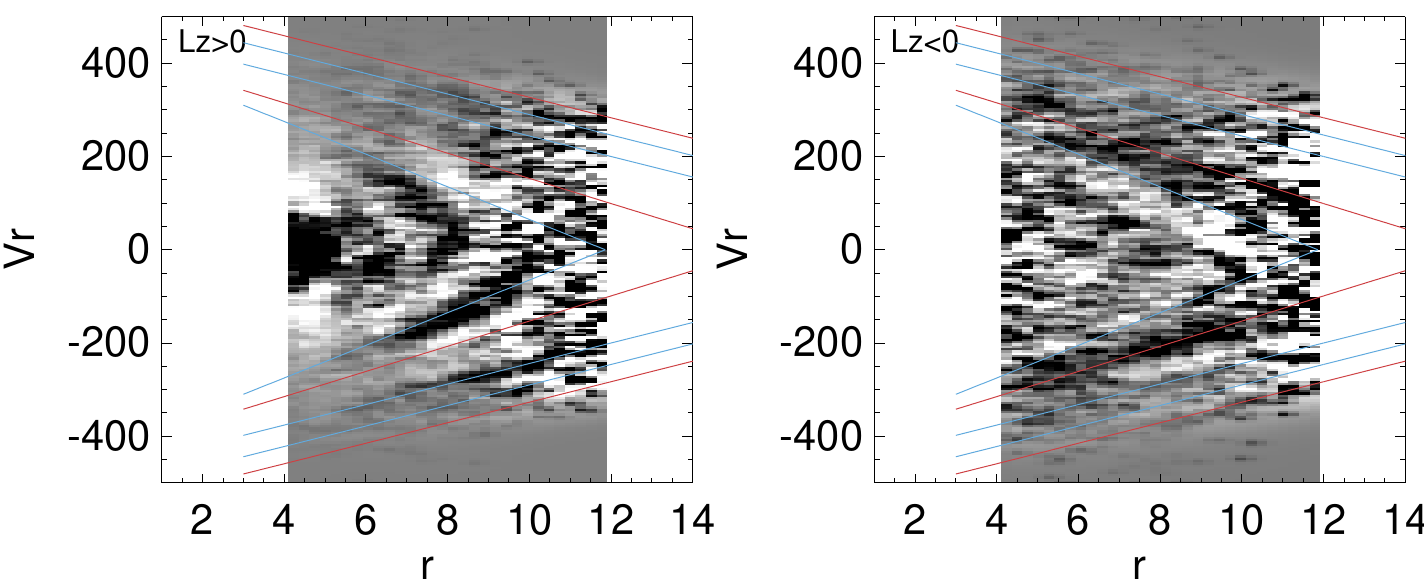}
  \caption[]{$v_r$ asymmetry in the detected phase-space folds. These background-subtracted density distributions are the same as in the bottom row of Figure~\ref{fig:phsp2}. The straight lines show the approximate behaviour of each chevron as gauged from the signal at $v_r<0$. Note that reflected around $v_r=0$ line, the model tracks tend to miss the overdensities in the $v_r>0$ portion of the phase-space. Also note that in $L_z<0$ view (right) chevron 1 exhibits a turn-around at $v_r\approx-50$ km s$^{-1}$ instead of 0 km s$^{-1}$.}
   \label{fig:phsp_asym}
\end{figure}

Figure~\ref{fig:vrr_max} presents a sketch of the phase-space properties of the GS/E debris. Following the notation introduced in the previous figures, the detected chevrons are shown in blue ($L_z>0$) and red ($L_z<0$) colour. Their behaviour is compared to the overall shape of the radial velocity distribution as given by thin grey lines marking the 0.5, 2, 5, 10, 20 and 99.5, 98, 95, 90, 80 percentiles of the $V_r$ velocity as a function of $r$. Dotted (solid) lines are for all stars with $|L_z|<0.5\times10^3$ (retrograde stars with $-0.5\times10^3<L_z<0$). The amplitude of the radial velocity variation mapped by dotted lines reaches maximum around 5 kpc from the Galactic centre, while the solid curves reach the peak at around 3 kpc. We surmise that the difference in the behaviour is due to the contribution of the in-situ halo (see the discussion in Section~\ref{sec:cosmo}), which is significantly lower for the retrograde selection. Overall, the evolution in the radial velocity amplitude matches the behaviour of the detected chevrons. This can be compared to the inferred radial velocity shape of the GS/E debris as mapped by the {\it Gaia} DR2 RR Lyrae reported in \citet{Iorio2021} and shown as a grey band. The radial velocity model of \citet{Iorio2021} is bi-modal, similar to that used in \citet{Lancaster2019} and \citet{Necib2019}. The $v_r$ distribution is approximated by two Gaussians whose separation in km s$^{-1}$ is allowed to vary with Galactocentric radius to mimic the phase-space density of a radial merger. The two grey bands in Figure~\ref{fig:vrr_max} show the inferred locations of the centers of the two Gaussians. As the stars on eccentric orbits approach their pericentres their radial velocity reaches its maximal value before dropping quickly at the turn-around radius. Thus for the GS/E debris the maximal $v_r$ amplitude is expected to be reached close to the overall debris pericentre. The trends in $v_r(r)$ shape reported here and by \citet{Iorio2021} are in good agreement. However, our retrograde selection (solid lines) imply that the maximum $v_r$ is reached closer to the Galactic centre and thus likely point at a slightly smaller pericentre ($r<3$ kpc).

\begin{figure}
    \centering
    \includegraphics{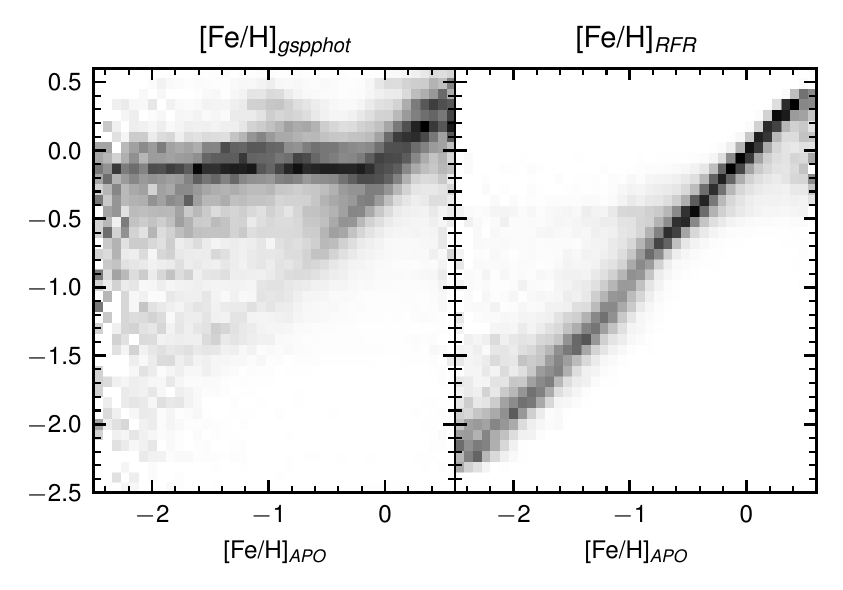}
    \caption{Comparison of [Fe/H] estimates derived from BP/RP spectra as published as part of the {\it Gaia} DR3 (gspphot, left) and computed here using the APOGEE DR17 labels (random forest regression applied to extinction corrected BP/RP coefficients, right).  Each panel shows the column normalized histogram (or essentially P(y$|$x) where x,y are the x,y-axes of the plot)}
    \label{fig:feh_measurements}
\end{figure}

Many of the chevrons detected are not symmetric with respect to the $v_r=0$ line. This is demonstrated in Figure~\ref{fig:phsp_asym} where the red (blue) curves mark the approximate locations of the $L_z>0$ ($L_z<0$) chevrons as gauged by the signal in the $v_r<0$ portion of the phase-space. These model chevron tracks are symmetric with respect to the $v_r=0$ line. However, on inspection of the Figure, it is clear that the symmetry is broken in the data. The clearest example is the behaviour of the chevron 1 in the $L_z<0$ sample (right panel) where it can be seen to turn around at $v_r\approx-50$ km s$^{-1}$. Such behaviour is only possible because chevrons are not orbits but are agglomerations of debris with different energies. Similarly, there is no clear counterpart to chevron 1 at $v_r>0$ in the $L_z>0$ sample (left panel). Also, the track for chevron 3 runs in between the two tentative detections at $v_r>0$. At $L_z<0$, the track for chevron 2 runs above the strongest signal at $v_r>0$. 
%This asymmetric behaviour of the phase-space folds may %be connected to the triaxial structure of the GS/E %debris \textbf{(check!)}.

%
\begin{figure*}
  \centering
  \includegraphics[width=0.99\textwidth]{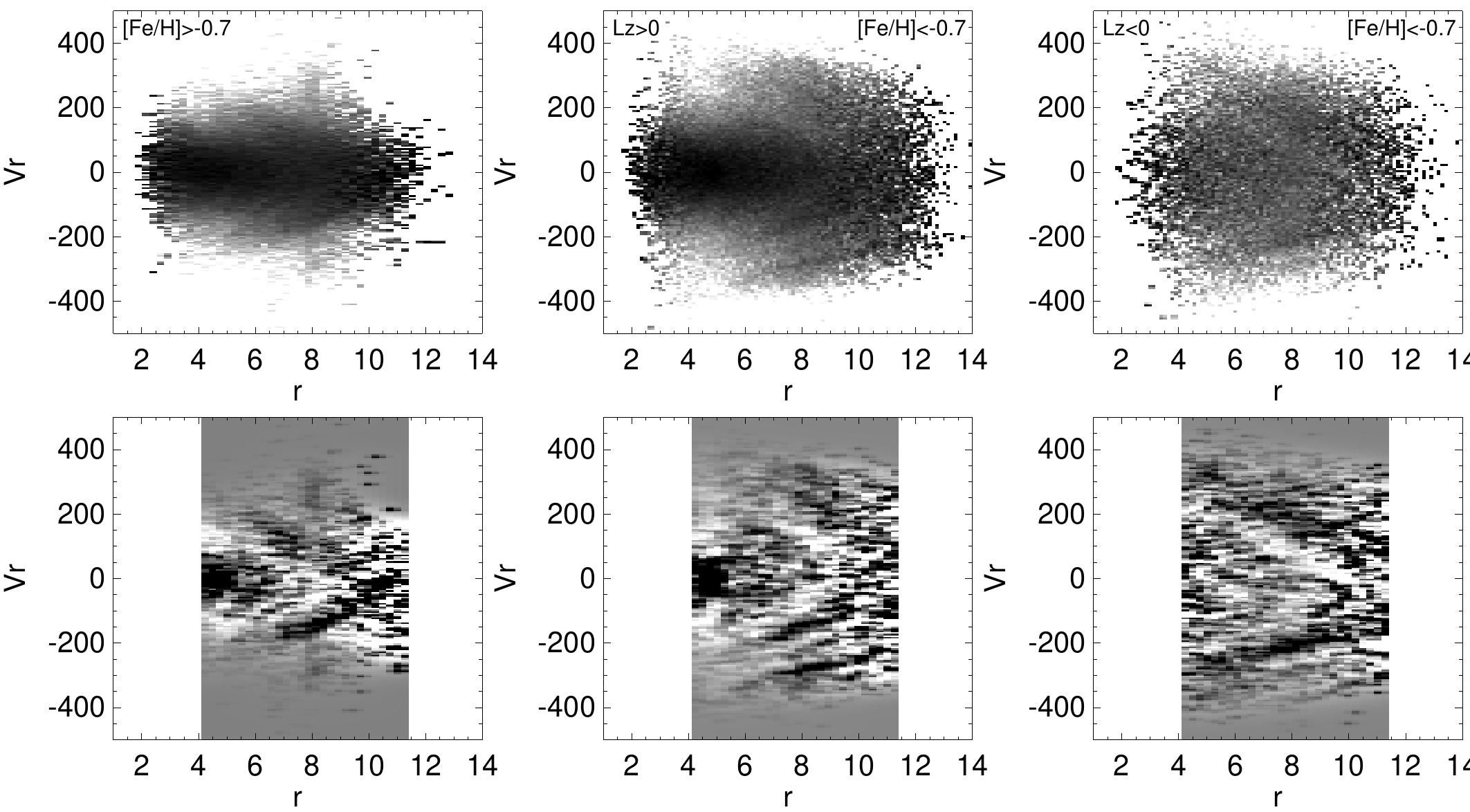}
  \caption[]{Metallicity dependence of the phase-space density in the Solar neighbourhood. {\it Left:} Stars with $|L_z<|0.75$ and [Fe/H]$>-0.7$ are used. This range of metallicity and angular momentum is dominated by the in-situ halo stars, as evidenced by much reduced range of $v_r$. Curiously, the in-situ phase-space distribution exhibits sharp density variations the most prominent of which appears to correspond to Chevron 1. {\it Middle:} Phase-space density for stars with $0<L_z<0.75$ and [Fe/H]$<-0.7$. Note that here Chevron 1 appears weaker and thinner compared to the view presented in Figure~\ref{fig:phsp2}. {\it Right:} Phase-space density for stars with $-0.75<L_z<0$ and [Fe/H]$<-0.7$ is largely unchanged compared to that shown in Figure~\ref{fig:phsp2}.}
   \label{fig:phsp_feh}
\end{figure*}

\subsection{Metallicity dependence of the local phase-space structure}
\label{sec:met}

In this Section we study the dependence of the detected phase-space susbtructure on metallicity. Because only $\sim$ 16\%  of the {\it Gaia} DR3 RVS sample have {\tt gspspec} metallicities reported (based on the RVS spectra), we instead use the metallicities derived from the BP/RP spectra. Unfortunately, the {\tt gspphot} metallicities, provided as part of the {\it Gaia} DR3, show significant biases (see left panel of Figure~\ref{fig:feh_measurements} for a comparison with the APOGEE DR17 abundances). This is likely caused by the overestimation of the dust extinction for some stars, which in turn affects temperature and [Fe/H] measurements. To mitigate these biases, we re-derive the metallicities from BP/RP spectra, using a data-driven approach calibrated to the APOGEE DR17 data \citep{apogee_dr17}. Specifically, we cross-match a subset of the {\it Gaia} DR3 sample with continuous mean-sampled BP/RP spectra ($\sim$ 220 million stars ) with the APOGEE DR17 measurements. We only use stars with accurate enough BP/RP reconstructions, i.e. {\tt bp\_chi\_squared <} 1.5 * {\tt bp\_degrees\_of\_freedom} and {\tt rp\_chi\_squared <} 1.5 * {\tt rp\_degrees\_of\_freedom}, leaving approximately 546,000 stars. Furthermore we exclude stars with extinction $E(B-V)>0.2$ from \citet{sfd98}, resulting in $\approx$344,000 stars.

\begin{figure*}
  \centering
  \includegraphics[width=0.99\textwidth]{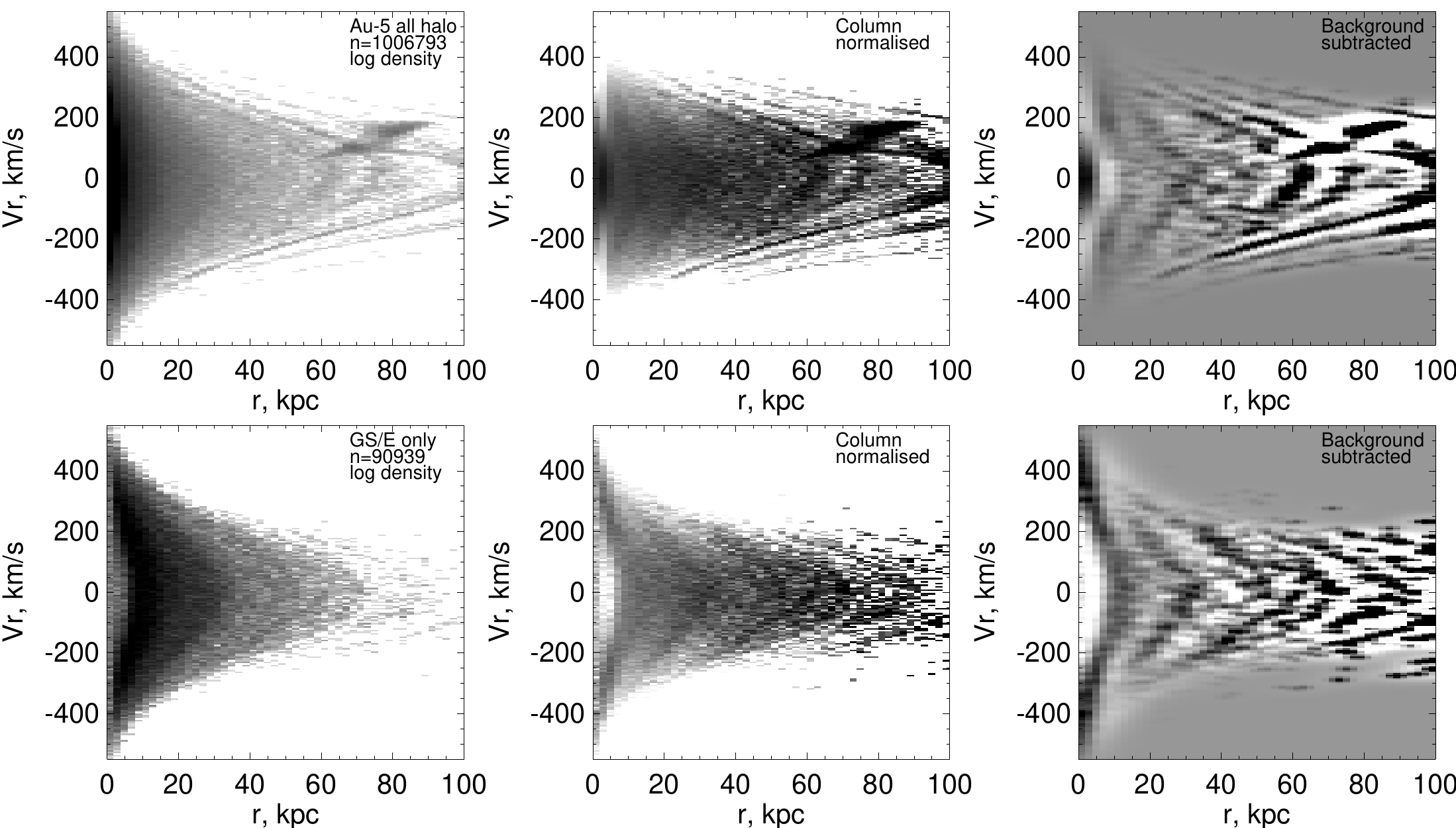}
  \caption[]{Phase-space $(v_r,r)$ in the Au-5 simulation. {\it Left:} Logarithm of stellar particle density. {\it Center:} Column-normalised density. {\it Right:} Background-subtracted density. {\it Top:} All particles classified as halo, i.e. with $|L_z|<0.7\times10^3$ including both the accreted and the in-situ components. {\it Bottom:} Only particles from the GS/E-like progenitor.}
   \label{fig:Auriga_large}
\end{figure*}
\begin{figure*}
  \centering
  \includegraphics[width=0.99\textwidth]{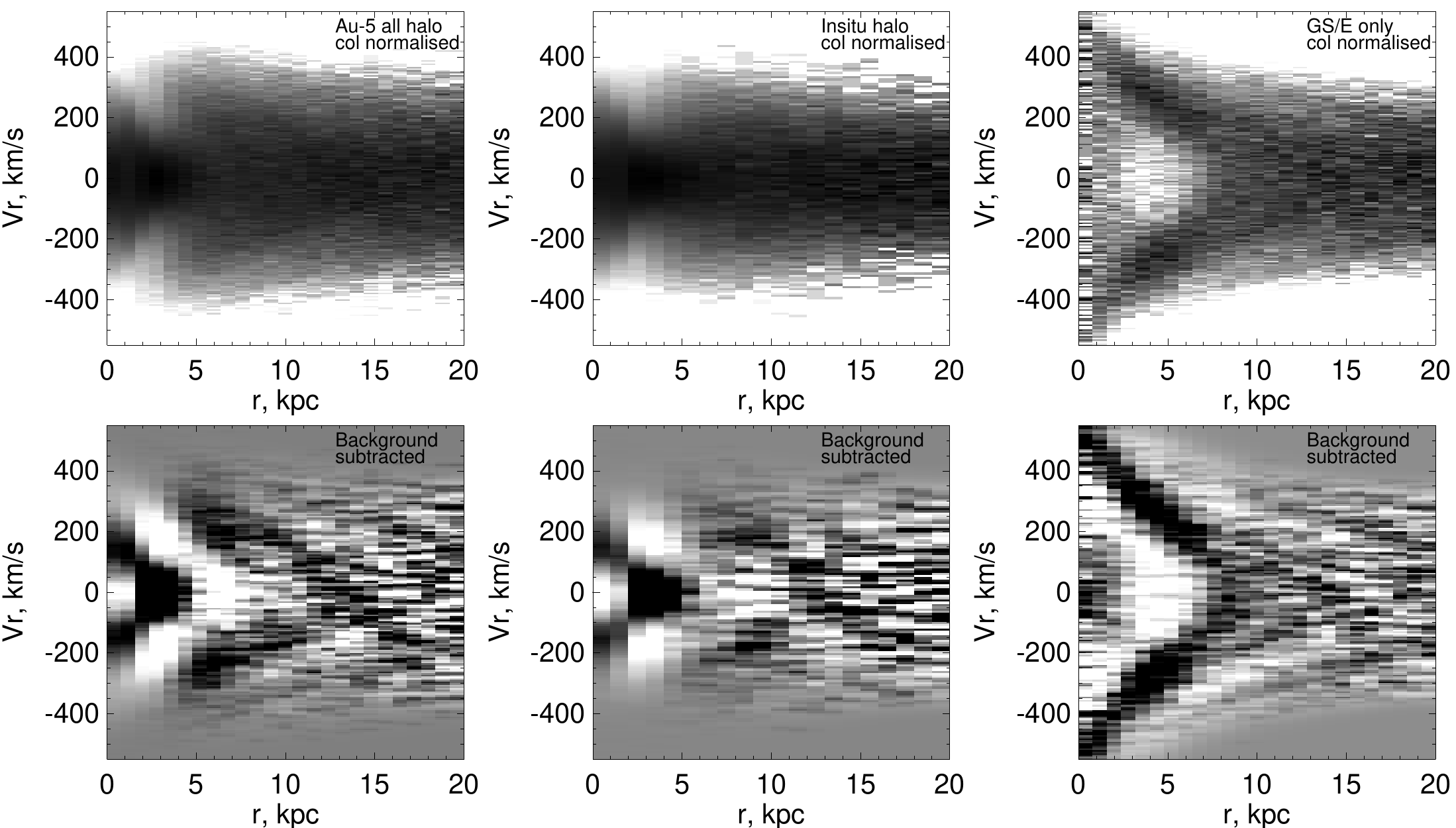}
  \caption[]{A zoom-in on the Solar neighbourhood portion of the $(v_r,r)$ phase-space in the Au-5 simulation. {\it Top:} Column normalised density. {\it Bottom:} Background-subtracted density. {\it Left:} All halo particles. {\it Center:} In-situ halo particles. {\it Right:} GS/E only particles.}
   \label{fig:Auriga_zoom}
\end{figure*}

We than use the random forest regression implemented in the \texttt{sklearn}\footnote{\url{https://scikit-learn.org}} package in order to map the array of 110 BP/RP spectral basis function coefficients onto [Fe/H]. In order to make the BP/RP coefficients independent from the brightness of stars, we divide each coefficient by the G-band flux of the star. To take into account the dust extinction, we also apply a linear extinction correction of the form  $X_0 = X -(C_0 + C_1 X_{\rm sub}) E(B-V)$, where $X_0 $ is the extinction corrected coefficient vector, $X$ is the original G-flux normalized BP/RP coefficient coefficient vector and $X_{\rm sub}$ is a subset of 10 leading coefficients from BP and from RP, and $C_0$ is a 110 element extinction vector, while $C_1$ is the 110$\times$10 extinction matrix. This extinction correction is only expected to be appropriate for small extinction values $E(B-V)\lesssim 0.5$ because at higher reddening values extinction stops being linear in the coefficient space. We use default parameters of the regressor and ignore the BP/RP coefficient uncertainties.
The right panel of the Figure~\ref{fig:feh_measurements} shows the resulting [Fe/H] values estimated with the random forest regression (RFR), plotted against the APOGEE measurements. Note that this shows the result of a cross-validation with 5 folds, in other words, the comparison should not be affected by over-fitting. The typical accuracy of the RFR (based on 16/84-th percentiles of the residuals) is $\sim 0.1$ dex for the sources with the magnitude distribution similar to the APOGEE sample. We note however that the stars in the {\it Gaia} DR3 RVS are typically brighter than the APOGEE DR17 targets. 

We use the spectrophotometric metallicities computed as discussed above to split the low-$|L_z|$ sample into two parts: the in-situ and the accreted halo. Note that with only [Fe/H] estimates in hand it is not possible to obtain a pure accreted selection, as the in-situ stars on halo-like orbits exists down to very low metallicities \citep[see][]{Aurora,Conroy2022,Myeong2022}. On the other hand, in the Solar neighbourhood, currently there is no strong evidence for GS/E members or any accreted stars with metallicities above [Fe/H]$=-0.7$. Therefore, we use [Fe/H]$=-0.7$ as the boundary between the accreted and the in-situ halo populations. 
Figure~\ref{fig:phsp_feh} shows the behaviour of the phase-space density for stars with [Fe/H]$>-0.7$ (in-situ sample, left panel) and [Fe/H]$<-0.7$ (accreted sample, middle and right panels). The range of variation of the in-situ radial velocities is lower compared to that of the accreted halo (i.e GS/E debris), in agreement with previous analysis \citep[][]{Belokurov2020}. However, the in-situ halo does not appear completely phase-mixed: its phase-space distribution shows strong density variations matching the location of Chevron 1. The fact the in-situ stars must have contributed to Chevron 1 in Figure~\ref{fig:phsp2} is confirmed in the middle panel of Figure~\ref{fig:phsp_feh}. Here, in the accreted sample, chevron 1's signal is visibly reduced.

\section{Comparison to Numerical Simulations}
\label{sec:sims}

\subsection{Cosmological Zoom-in Simulations}
\label{sec:cosmo}

The Auriga cosmological magneto-hydrodynamical simulations \citep{Grand2017} consist of a sample of 30 MW-mass haloes simulated using the zoom-in technique \citep{Jenkins2013}. The haloes are selected from a parent dark-matter-only box\footnote{The parent box is the DMO counterpart of the largest box from the EAGLE project \citep{Schaye2015}} of size $(100\,{\rm Mpc})^3$, and have been constrained to be isolated and in the mass range $M_{200,c}=(1-2)\times10^{12}$ M$_{\odot}$ at $z=0$.
The simulations adopt cosmological parameters based on \citet{Planck2014} and were performed using the $N$-body magneto-hydrodynamical code Arepo \citep{springel2010}. In summary, the galaxy formation model includes homogeneous UV background radiation, gas cooling, star formation, stellar evolution feedback, black-hole accretion and AGN feedback. Analysis of this section is based on the 30 haloes simulated at the ``L4 resolution'' with dark matter and baryonic resolution elements of mass $\sim 3\times 10^5 M_{\odot}$ and $5\times10^4 M_{\odot}$, respectively. 

\begin{figure}
  \centering
  \includegraphics[width=0.49\textwidth]{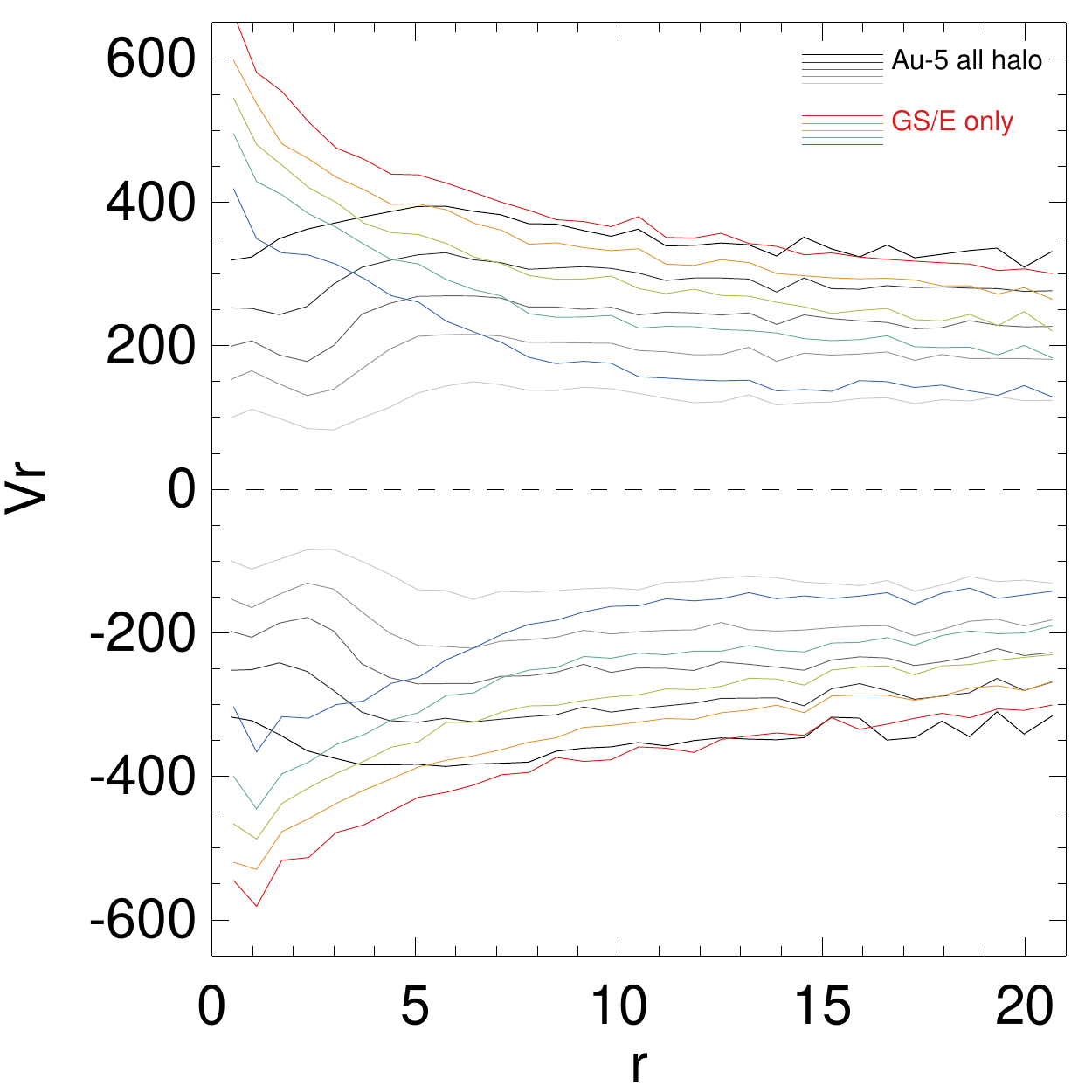}
  \caption[]{Similar to Figure~\ref{fig:vrr_max} for the cosmological simulation Auriga-5. The change in shape of the phase-space density captured by the percentiles of the $v_r(r)$ distribution. Grey lines correspond to all halo particles, coloured lines show the behaviour of the GS/E particles. While the amplitude of the GS/E of the $v_r(r)$ variation keeps increasing towards small radii, when all halo particles are considered a turnover at $\sim5$ kpc appears. The distribution changes shape because in the inner Galaxy the density is dominated by the in-situ halo.}
   \label{fig:Auriga_vrr_max}
\end{figure}

\citet{Fattahi2019} created mock observations of the redshift $z=0$ stellar halos of the Milky Way analogs in the Auriga suite. Using a Gaussian Mixture Model \citep[see][]{Belokurov2018} they identified the Auriga hosts whose stellar halos (around the Solar radius) contained a highly elongated feature in the space of radial and azimuthal velocities. Requiring a high radial anisotropy and a high fractional contribution of the GS/E-like debris to the halo's stellar mass, \citet{Fattahi2019} selected a group of hosts whose stellar halos at Solar radius best resembled the observed local stellar halo of the Milky Way. While approximately one third of all Auriga hosts satisfied loosely the above conditions, four systems in particular, namely Au-5, Au-9, Au-10, Au-18, stood out in terms of their high radial anisotropy and the dominance of the GS/E-like debris. The stellar masses of the GS/E-like progenitor galaxy in these are between 1 and 4$\times10^9M_{\odot}$ and the accretion look-back times are between 7 and 11 Gyr. Of these four, the GS/E merger in Au-5 is the most massive and therefore contains the largest number of star particles, some three to four times more than the other three simulations. As the GS/E debris phase-mixes and thins out, the number of particles in the simulation plays a crucial role in our ability to resolve the folds \citep[see][for a detailed discussion of the role the sampling of the phase-space density plays in detecting the amount of mixedness]{Leandro2019}. We therefore focus on Au-5, as it provides just enough of a particle resolution to study the survival of the phase-space chevrons of a GS/E's numerical counterpart.

Figure~\ref{fig:Auriga_large} gives the $(v_r,r)$ phase-space density of the Au-5 halo with 200$\times$50 pixels. Left to right, the columns show the logarithm of the stellar particle density, the column-normalised density and the background subtracted density. As above, the background is obtained by simply convolving the density distribution with a Gaussian (in this case with a FWHM=7 pixels). The top row shows all of the halo particles, including both the accreted and the in-situ components. These are selected by requiring $|L_z|<0.7\times10^3$ similar to the cuts applied to the data in the previous section. The bottom row shows only particles from the GS/E-like progenitor. As the top row demonstrates, the phase-space of Au-5's halo contains a large number of nested folds. However, on the inspection of the bottom row, it is apparent that some of these belong to other accretion events.  The bottom row of Figure~\ref{fig:Auriga_large} demonstrates that phase-space folds left behind by a GS/E-like event survive to the present day across a wide range of Galactocentric distances.

\begin{figure*}
  \centering
  \includegraphics{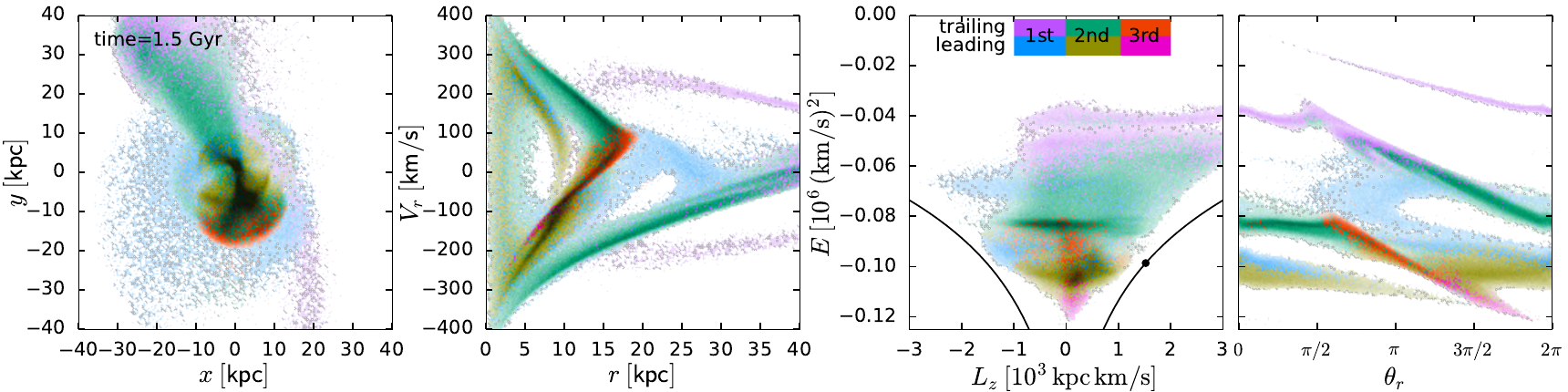}
  \includegraphics{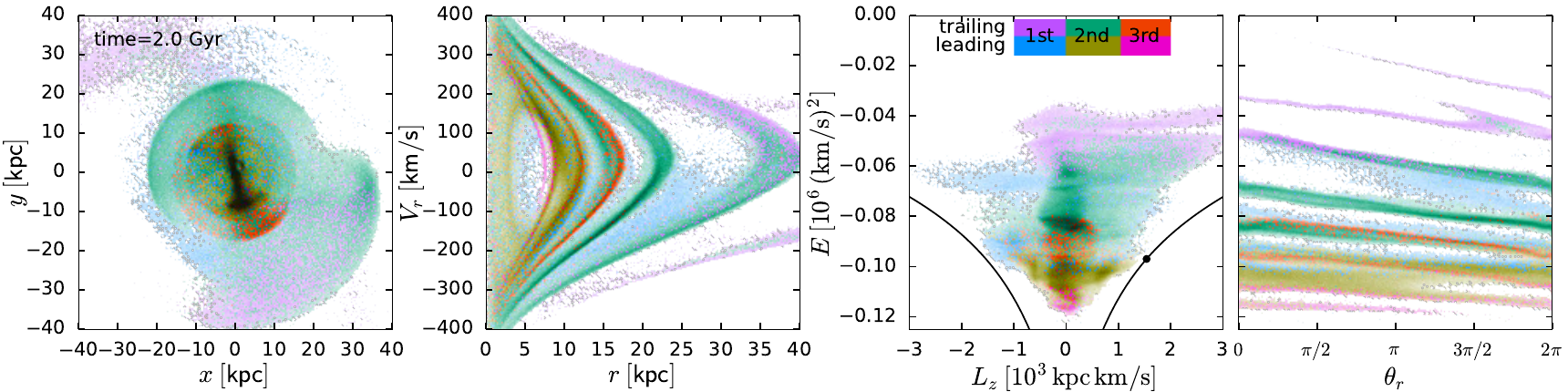}
  \includegraphics{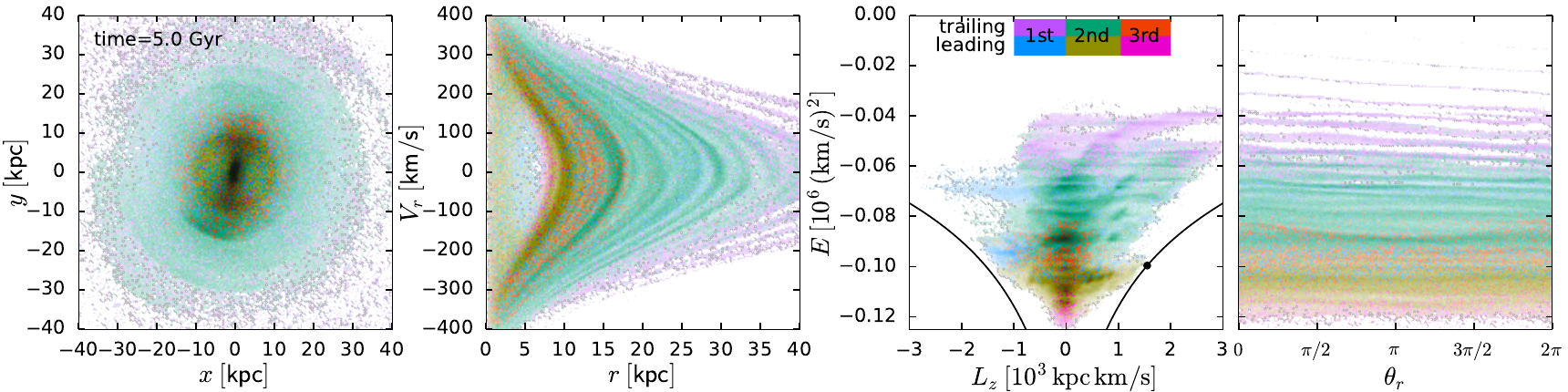}
  \includegraphics{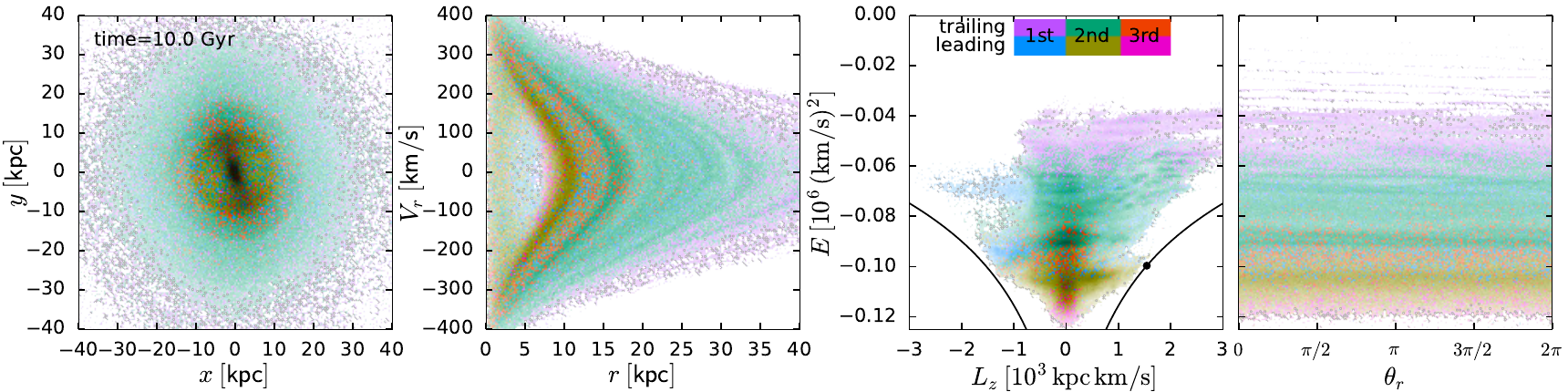}
  \caption{
  Tailored $N$-body merger simulation in different views. Top to bottom rows show the snapshots at different moments of time: 1.5~Gyr (shortly after the last pericentre passage of the satellite leading to its full disruption), 2, 5 and 10~Gyr. Leftmost column shows the spatial distribution of the debris, second column shows the $r$--$v_r$ phase-space, third panel presents the $E$--$L_z$ space, and the rightmost one shows $E$ vs.\ the radial phase angle $\theta_r$. Particles are coloured by their stripping episode and the location in the leading or trailing arms at the time of unbinding from the satellite. In the third column, black lines delineate the angular momentum of a circular orbit, and black dot marks the fiducial Solar location in this space.
  }
  \label{fig:mergersim_global}
\end{figure*}

Figure~\ref{fig:Auriga_zoom} presents a zoomed-in view of the Au-5's phase-space density around the Solar radius, for a direct comparison with the observed behaviour discussed in the previous Section. From left to right, the columns show all halo particles, the in-situ halo particles and the GS/E only particles. The column-normalised (background-subtracted) density is given in the top (bottom) row of the Figure. The GS/E distribution contains a number of chevrons in the Solar neighbourhood, i.e at $5<r$(kpc)$<15$ (bottom right). Note however, that when all halo particles are considered, the strength of some of these is reduced (bottom left). This is partly due to the contribution of the in-situ halo (bottom centre). While the orbits of the GS/E particles reach as close as $\sim1$ kpc from the Galactic centre, at these small Galactocentric distances they are subdominant, being overwhelmed by the in-situ halo (compare the left and the right panels in the top row of the Figure). This is illustrated in Figure~\ref{fig:Auriga_vrr_max} which serves as a companion to Figure~\ref{fig:vrr_max} and shows the overall shape of the $(v_r,r)$ phase-space as indicated by the percentiles of the radial velocity distribution in bins of radius. Similarly to Figure~\ref{fig:vrr_max}, the grey lines peak around $r\approx5$ kpc. However, as clear from the behaviour of the coloured curves showing the $v_r(r)$ distribution trends for the GS/E, this radius can not be interpreted as the radius of the maximal amplitude of the GS/E radial velocity. In Au-5, the turnover in the all-halo curves at $r\approx5$~kpc is driven by the in-situ halo contribution.

\subsection{Tailored merger simulations}
\label{sec:mergersim}

\begin{figure*}
  \centering
  \includegraphics{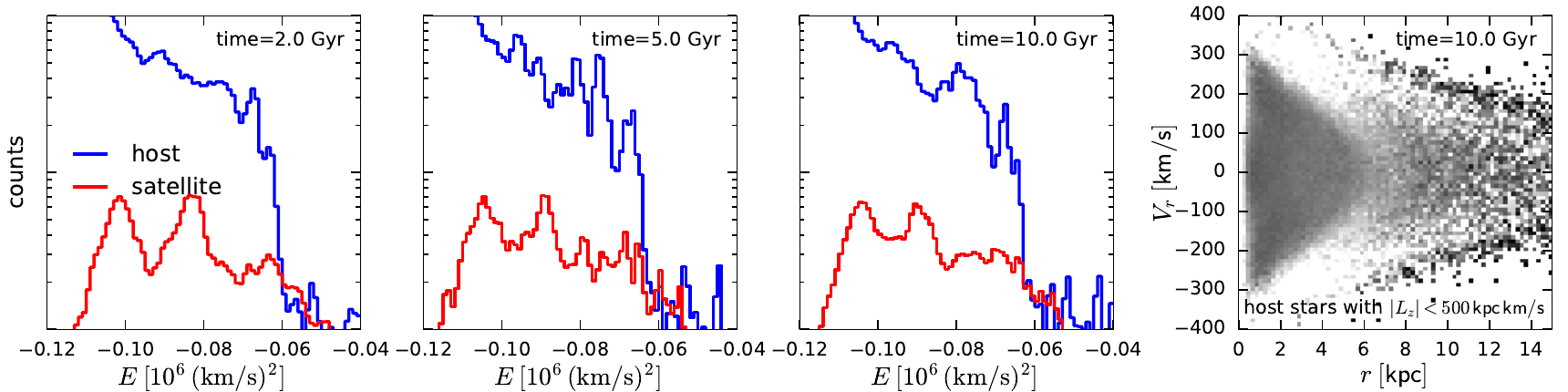}
  \caption{
  First three panels: energy $E$ distribution of host (blue) and satellite (red) particles in the tailored merger simulation at different times (one-dimensional projection of the third column in Figure~\ref{fig:mergersim_global}). The two most prominent bumps in the accreted population at $E=-0.1$ and $-0.08$ correspond to the leading and trailing arms (olive and green in the previous figure), while the smaller-scale clumps at higher energy appear in both \textit{in situ} and accreted populations within a few Gyr after the merger and then gradually dissolve. Right panel: unsharp-masked distribution of low-angular-momentum ($|L_z| < 500\,\mathrm{kpc\,km\,s}^{-1}$) host stars at the end of the simulation, displaying a chevron-like feature.
  }
  \label{fig:mergersim_insitu}
\end{figure*}

\begin{figure*}
  \centering
  \includegraphics{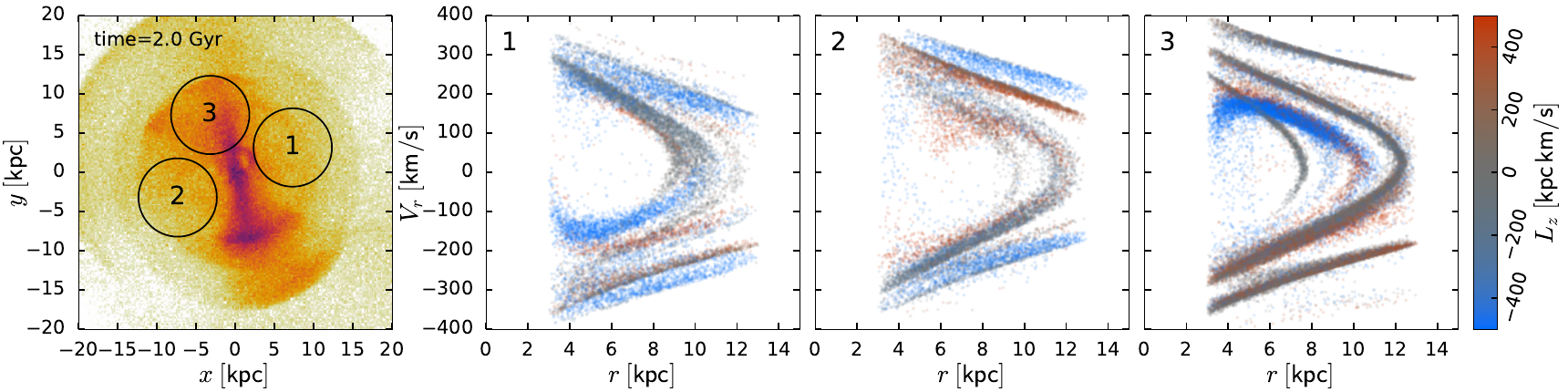}
  \includegraphics{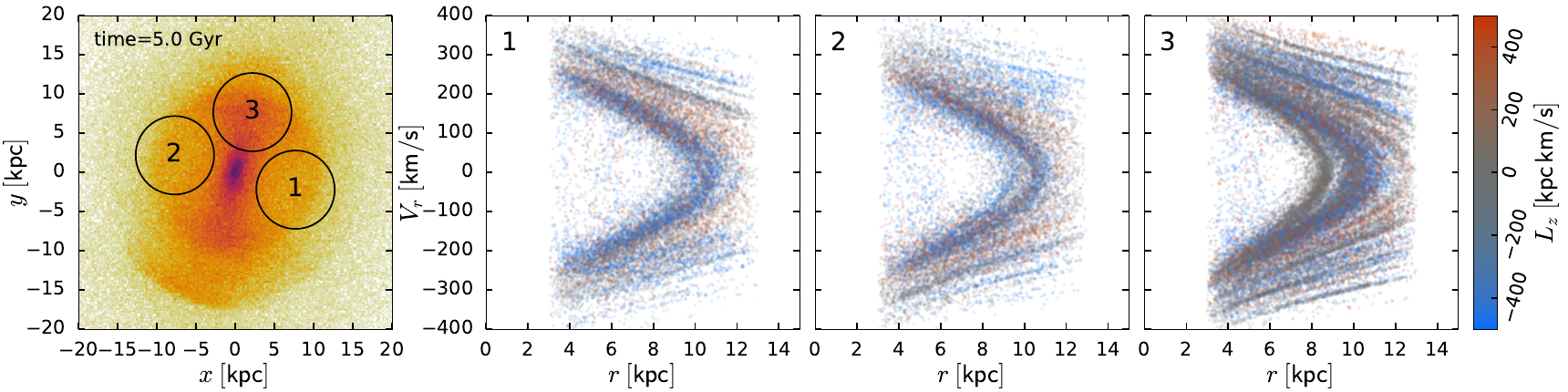}
  \includegraphics{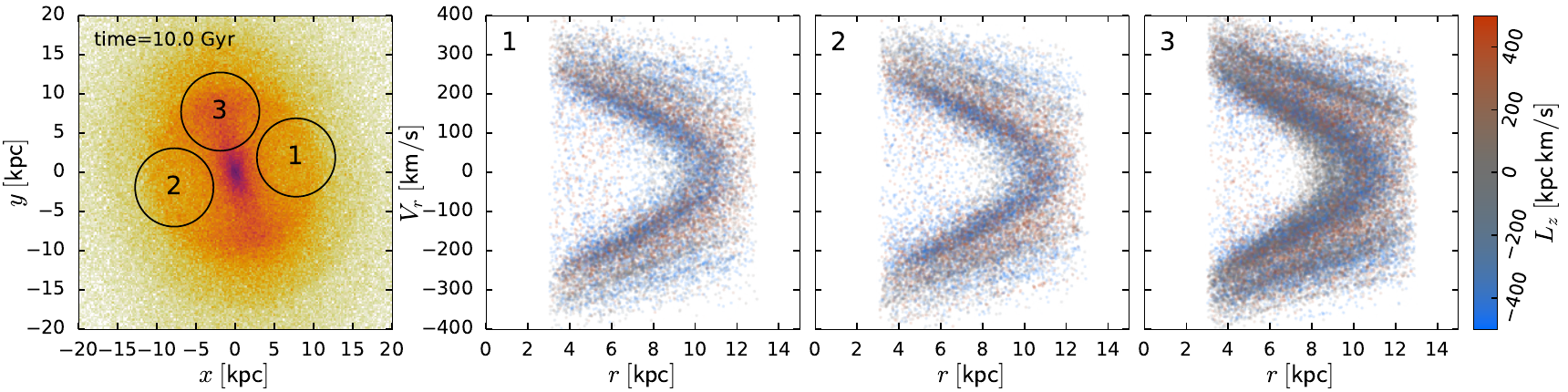}
  \caption{
  Snapshots from the tailored merger simulation (same as the second to fourth rows in Figure~\ref{fig:mergersim_global}), focusing on the distribution of merger debris in smaller spatial regions. Left column shows the distribution of merger debris, indicating the location of three fiducial regions (spheres of radius 5 kpc centered at 8 kpc from origin). Regions 1 and 2 are positioned along the minor axis of the merger debris in the equatorial plane, similarly to the Solar position w.r.t.\ the GS/E population in our Galaxy, whereas region 3 is located along the major axis of the debris cloud. Remaining columns show the $r$--$v_r$ space in these regions, with particles coloured by their angular momentum $L_z$. Although the evolution proceeds qualitatively in the same way across the entire galaxy, with individual chevrons winding up more tightly as time goes on, the difference between spatial regions can be quite significant.
  }
  \label{fig:mergersim_local}
\end{figure*}

In order to have a greater control on the properties of the merger remnant, we ran a number of dedicated $N$-body simulations of mergers between a Milky Way-like host galaxy and a GS/E progenitor (satellite). We broadly follow the simulation setup described in \citet{Naidu2021}: the mass ratio between the host and the satellite is 1:2.5, its initial orbit is moderately eccentric (orbital circularity $\eta\equiv L/L_\mathrm{circ} = 0.5$), and the orbital angular momentum is tilted by $30^\circ$ from the angular momentum of the host disc; however, unlike the \cite{Naidu2021} study, our satellite orbit is prograde. Both galaxies are initially set up in equilibrium, with $(1+4)\times10^6$ stellar and dark matter particles in the host and $(0.5+2)\times10^6$ particles in the satellite. A detailed description and analysis of our simulation suite will be presented elsewhere, and here we focus on one particular model, which qualitatively reproduces many of the observed properties of the GS/E population, but does not necessarily match the Milky Way in detail. In particular, the total mass and therefore the energy scale of the merger remnant are somewhat lower than in the fiducial Milky Way potential used in Section~\ref{sec:energy}. In this simulation, the satellite barely completes three pericentre passages before being fully disrupted, and as discussed in \citet{Naidu2021} and \citealt{Vasiliev2022}, the orbit of such a massive satellite quickly radializes, so that the final angular momentum of the debris is close to zero. 

Figure~\ref{fig:mergersim_global} shows a sequence of four snapshots at different times, starting from the moment just after the disruption and ending roughly at the present day. We separate the stellar particles of the satellite into three different stripping episodes, with the second one being the most dramatic, and within each episode, by their location in the leading or trailing arms (based on the energy difference between the satellite centre and the particle at the moment of its unbinding). The least bound populations (in the first and the second trailing arms, magenta and green) have on average slightly positive $L_z$, inheriting it from the satellite's orbital angular momentum, and the more tightly bound debris are mostly located closer to $L_z=0$. However, some of the stars from the first leading arm (cyan) occupy the region with sufficiently negative $L_z$ to be associated with the Sequoia \citep{Myeong2019} and Thamnos \citep{Koppelman2019b} populations in the Milky Way. Given that the satellite orbit was initially prograde w.r.t.\ the host disc, it may appear surprising that the first to be stripped debris end up in the retrograde region. However, we stress that the host galaxy also moves significantly during this high-mass-ratio merger, so the angular momentum w.r.t.\ the host centre is not conserved even for particles that are no longer bound to the satellite. It is therefore plausible that these retrograde populations may come from the same progenitor galaxy as GS/E itself, as advocated by \citet{Koppelman2020} and \citet{Amarante2022}. In addition, the angular momentum of the host galaxy disc after the merger may continue to precess \citep{Dillamore2022, Dodge2022}, further complicating the interpretation of the $E$--$L_z$ distribution of the debris.

Particles stripped in each episode have a wide range of energies and orbital periods, leading to the phase mixing and winding up of individual chevrons in the $r$--$v_r$ phase-space and a corresponding stretching and flattening of ridges in the $E$--$\theta_r$ space (see Section~2 in \citealt{DongPaez2022} for an in-depth discussion). Each chevron in the $r$--$v_r$ space corresponds to a continuous segment of particles stretching from $\theta_r=0$ (pericentre) through $\theta_r=\pi$ (apocentre) to $\theta_r=2\pi$ (next pericentre). As the fourth column of Figure~\ref{fig:mergersim_global} illustrates, particles belonging to the same chevron have a gradient of energy vs. $\theta_r$, with the less bound particles having longer radial orbital periods and therefore moving slower to the right in the $\theta_r$ direction. Upon crossing the right boundary $\theta_r=2\pi$, particles reappear on the left boundary and continue moving to the right with a constant speed, increasing the number of chevrons with time.
Particles that arrive to their apocentres later have higher energies and therefore turn around at larger radii. This creates the prominent asymmetry of the outermost chevrons w.r.t.\ the sign of $v_r$, with the maximum radius corresponding to positive $v_r$ and therefore moving outward with time. Eventually, the gradients in the $E$--$\theta_r$ space decrease and the individual chevrons become more and more monoenergetic, but the distance in energy or radius between them also decreases, making it difficult to separate out individual wraps. On the other hand, the ``super-chevrons'' (agglomerations of individual folds) corresponding to the same stripping episode and arm (olive and green) remain discernible even after 10~Gyr, and their spacing is primarily determined by the difference in energy between the leading and the trailing arms of the main (second) stripping episode, which itself encodes information about the progenitor mass and structure. The apocentre radii of the super-chevrons correspond to bumps and breaks in the density profile of the merger debris \citep{Deason2018}.

We also note that the distribution of particles in energy space becomes more wrinkled within a few Gyr after the merger (third row in Figure~\ref{fig:mergersim_global}). This is likely caused by global modes in a self-gravitating merger remnant, where particles satisfying particular resonance conditions end up evolving in a similar way and reinforcing perturbations \citep[e.g.,][]{Weinberg1989}. However, these small-scale structures are eventually smoothed out and dissipate (fourth row), except perhaps in the outermost parts of the remnant. 
Figure~\ref{fig:mergersim_insitu} shows that these wrinkles in the one-dimensional distribution of particles in energy are present both in the accreted and \textit{in situ} populations, and that the $r$--$v_r$ phase-space distribution of the host stars also has at least one chevron-shaped feature created during the merger (corresponding to the \textit{Splash} population in the Milky Way).

Figure~\ref{fig:mergersim_local} illustrates the spatial variations in the $r$--$v_r$ phase-space distribution at different times. Shown are three spatial regions of radius 5 kpc, all centered at 8 kpc from origin, but at different azimuthal positions; of these, the first two are similar to the actual Solar location relative to the GS/E population in the Milky Way (Figure~3 in \citealt{Iorio2019}). At all times, the chevron configuration varies with the selected location and is different above and below $v_r=0$. Although the detailed structure of chevrons depends on the local spatial region, in general a few individual folds remain visible after 10~Gyr, especially when coloured by $L_z$. The variation in $L_z$ between chevrons is inherited from the most dynamic stages of the merger, where both the host and the satellite move relative to each other, and debris stripped at different times and energies end up having different values of $L_z$ after the dust settles.

\section{Discussion and Conclusions}
\label{sec:conc}

We have used {\it Gaia} DR3 RVS data to study small-scale sub-structure in the local portion of the tidal debris from the last significant merger, known as GS/E. The {\it Gaia} DR3 RVS dataset increases the number of stars with the complete 6D phase-space information by an order of magnitude compared to other currently available spectroscopic surveys such as SDSS, RAVE, LAMOST, GALAH or APOGEE. For example, applying selection criteria similar to those used in our analysis (e.g. the quality of the astrometric distance measurement and angular momentum) to APOGEE DR17 and LAMOST yields $\approx10,000$ and $\approx30,000$ halo stars in each of these respectively compared to $\approx260,000$ analysed here.  It is this ramp-up in the resolution that enables the study of the small-scale halo density variations reported here. 

Armed with the {\it Gaia} DR3 RVS data, we are able to discover a large number of previously undetected sub-structures in the density distribution of the local stellar halo (identified here with a simple angular momentum selection), both in the phase-space and the integrals-of-motion space. In the energy and angular momentum $(E, L_z)$ space, the GS/E debris is revealed to have an elongated and tilted, avocado-like shape where stars with higher energy tend to have $L_z>0$. The largest number of stars are at low energies, in the pit of the avocado. The $E$ distribution is also rather wrinkly with several overdensities and depletions, visible most clearly when halo stars with extreme (either prograde or retrograde) $|L_z|$ are considered (see Section~\ref{sec:energy} and Figure~\ref{fig:elz}). Similarly, in the $(v_r,r)$ phase-space the local stellar halo density is not smooth. It shows a network of thin, nearly linear bands or chevrons that are over-dense compared to the mean background. These chevrons show different strength depending on whether stars with $L_z>0$ or $L_z<0$ are considered (Section~\ref{sec:phase-space} and Figure~\ref{fig:phsp2}). Part of the difference is due to the in-situ halo contamination which is a strong function of the angular momentum. At least 5 distinct chevrons are recognisable in the $v_r<0$ portion of the phase-space, however some of these may be further resolvable into narrower components (Figure~\ref{fig:phsp3}).

To verify the in-situ halo contribution we estimate stellar metallicities by training a random forest regressor on the {\it Gaia} DR3 BP RP spectra labelled using APOGEE DR17 abundances (Section~\ref{sec:met}). Figure~\ref{fig:phsp_feh} confirms that a large fraction of low-energy low-$L_z$ stars are metal-rich, with [Fe/H]$>-0.7$ and thus likely born in the Milky Way proper. The phase-space density of the in-situ halo exhibits several sharp changes including a chevron-like feature overlapping with chevron 1. The in-situ halo contribution needs to be taken into account when estimating the pericentre of the GS/E debris: as judged by the retrograde members only it must be within $r\approx3$ kpc of the Galactic centre (Figure~\ref{fig:vrr_max}). The most prominent chevrons 1 and 2 have energies similar to those of the main peaks of the energy distributions (compare Figure~\ref{fig:phsp3} and Figure~\ref{fig:elz}). These strong overdensities ought to influence the stellar halo's radial density profile likely causing an additional break around $10<r$(kpc)$<15$ as proposed in \citet{Naidu2021} and  \citet{Han2022}.

As far as the interpretation of the discovered halo sub-structures is concerned, both the energy wrinkles and the phase-space chevrons are likely to be the consequence of phase-space mixing. In the Solar neighbourhood, i.e. away from both the pericentre and the apocentre, only the quasi-linear portions of chevron are accessible and, indeed, this is how many of these structures appear in the {\it Gaia} data. Phase-mixing is not supposed to alter the $E$ distribution. However, as a result of the mixing, stellar energies end up correlating strongly with the phase-space coordinates, i.e. stellar positions and velocities.  Due to the ensued energy sorting, only stars with specific orbital frequencies will be at the right phase to be observed in a small region in the configuration space (e.g. a volume around the Solar neighbourhood) today. This would lead to a series of depletions in the energy distribution of the tidal debris which, as a result, would appear wrinkled. Similarly, even if due to phase-mixing the number and the size of chevrons evolved to be such that at the present day individual chevrons are no longer discernible, limiting the sample to a small region around the Sun would pick out a subset of chevrons, making the phase-space more striated.

The above discussion describes the situation when a single clump of stars is deposited by the disrupting satellite into the host's potential to phase-mix. Realistically, however, even in the case of an explosive dissolution of rapidly radializing satellite -- such as that expected for the GS/E progenitor -- multiple stripping episodes are predicted, each resulting in at least two clumps of stars with distinct energies (and often $L_z$) corresponding to the leading and the trailing tidal debris. As the satellite sinks in the host's potential, the energy of each deposited clump will be different, thus making the final energy distribution clumpy (Section~\ref{sec:mergersim}). This may lead to several complications in the future analysis of the {\it Gaia} data. First, any modelling techniques that rely on the assumption of a smooth stellar distribution function \citep[e.g.][]{Leonard1990} may struggle given how wrinkly the observed DF is \citep[see][]{Grand2019}. Furthermore, linking overdensities in $(E, L_z)$ space with individual accretion events, following ideas of e.g. \citet{Helmi2000}, may easily be fraught with danger because a single event can produce multiple clumps.

As the satellite loses mass, the energy spread of the stripped stars diminishes, thus the energy distribution of the debris from a sinking and disrupting dwarf galaxy should appear asymmetric, peaking at low $E$ as observed here (see Section~\ref{sec:energy}). Stars in each energy overdensity left behind will phase-mix and create their own set of chevrons, which will evolve and merge, creating thicker super-chevrons (see Section~\ref{sec:mergersim}). It appears therefore that the complexity in the $(E, L_z)$ distribution of the GS/E tidal debris makes the use of the detected chevron patterns for the timing of the merger event rather difficult. Nonetheless, the first comparison with both tailored and Cosmological zoom-in simulations are reassuring, as most of the salient features uncovered in the {\it Gaia} data have numerical counterparts in the models analysed. As demonstrated by an example from the Auriga Cosmological zoom-in suite, phase-space chevrons left behind by a merger similar to GS/E can remain detectable in the host for many Gyrs (Section~\ref{sec:cosmo}). Snapshots of our bespoke simulations show a clear prograde tilt of the $(E, L_z)$  cloud as well the asymmetric, bottom-heavy energy distribution. Thanks to the high resolution of our tailored N-body simulation of a GS/E-like merger, new details of both the disruption and the relaxation processes emerge. For example, the energy distribution gets wrinkled not only due to the bursty stripping history but also because of the emergence of what appears to be global and rapid DM density oscillations, ``sloshing'' through the entire host soon after the satellite dissolves. These ripples (Figure~\ref{fig:mergersim_insitu}) flatten out with time, and at the present day, inhomogeneities in the energy distribution are dominated by the individual mass loss events and the leading-trailing energy bimodality. The persistence of these feature over many Gyr should help trace and reconstruct the sinking of the GS/E progenitor as its orbit radialized. 

The simulated local Solar neighbourhood phase-space distribution bears many a resemblance to the {\it Gaia} DR3 RVS observations. For example, the chevron pattern varies noticeable with both $v_r$ and $L_z$ (Figure~\ref{fig:mergersim_local}). These $L_z$-related variations are at least in part due to the complexity of the initial conditions with which the debris are deposited into the host. The differences in the chevron properties at positive and negative $v_r$ are possibly due to either non-zero angular momentum of the stars considered and/or the asphericity of the host potential. Both factors would ensure that once the stars have passed through the Solar neighbourhood, they are likely to miss this relatively small volume on the way back after the turn-around. Our tailored simulations convincingly demonstrate that the amount of small-scale substructure in the phase-space does not only depend on the time since the beginning of the merger but also on the observer's position inside the debris cloud. While many narrow chevrons are detectable even at 10 Gyr, the late-time phase-space density is dominated by super-chevrons. These phase-space folds observable as e.g. $(v_r, r)$ chevrons today are a powerful tool for constraining the MW's gravitational potential similar to other phenomena that arise due to long-term phase-mixing, such as stellar streams in the halo or ``snails'' (phase spirals) in the disk.

Note that phase-space folds have the advantage of covering a larger volume in the Galaxy compared to a typical stream. This can be exploited to map out the global properties of the Milky Way's mass distribution in a novel and independent way. A demonstration of this idea is presented in Dey et al (2022) who detect multiple phase-space chevrons in the halo of the Andromeda galaxy. Enjoying the wide and dense spectroscopic coverage provided by the recently launched DESI survey they show that the Andromeda's Giant Stellar Stream is a part of a complex system of tidal debris from a massive and recent accretion event. Given the merger's relatively early stage, in contrast to that of the GS/E, the chevrons identified by Dey et al (2022) are fewer, more pronounced and are set further apart from each other in the phase space compared to the sub-structures discovered here. Dey et al (2022) take advantage of the panoramic view of the entire M31 halo and use the  phase-space folds to constrain the M31's potential out to $>$100 kpc from its centre. 

Additionally, being as fragile as streams, the phase-space folds will blur and dissolve in reaction to time-varying potential perturbations \citep[also see][for a discussion of additional effects that can lead to the dissolution of the phase-space sub-structure]{Leandro2019evo}. Interactions with a passing mass, such as a sub-halo flying by or a rotating bar, would alter the affected stars' orbital frequencies and thus destroy the coherence of the chevrons. Therefore, the continued existence of these features implies a relatively quiet evolution of our Galaxy since the GS/E merger, although further work is needed to make more quantitative statements. We hope to address the response of the phase-space folds to satellites such as the Sgr dwarf in a forthcoming publication (see Davies et al 2022). 

\section*{Acknowledgments}

The authors are grateful to Kathryn Johnston, Nicol\'as Garavito-Camargo, Emily Cunningham, Adrian Price-Whelan, David Spergel, Jason Hunt, Giuliano Iorio, Leandro Beraldo e Silva, Jason Sanders, David Hogg, Iulia Simion, Melissa Ness and the members of the Cambridge Streams and the CCA Dynamics groups for many illuminating discussions that helped to improve the quality of this work. VB thanks Kohei Hattori for enlightening conversations about phase-mixing at the Santa Barbara Gaia Sprint held at the KITP UCSB in Spring 2019. This project was developed in part at the Gaia F\^ete, hosted by the Flatiron Institute Center for Computational Astrophysics in 2022 June.

This research made use of data from the European Space Agency mission Gaia
(\url{http://www.cosmos.esa.int/gaia}), processed by the Gaia Data
Processing and Analysis Consortium (DPAC,
\url{http://www.cosmos.esa.int/web/gaia/dpac/consortium}). Funding for the
DPAC has been provided by national institutions, in particular the
institutions participating in the Gaia Multilateral Agreement. This
paper made used of the Whole Sky Database (wsdb) created by Sergey
Koposov and maintained at the Institute of Astronomy, Cambridge with
financial support from the Science \& Technology Facilities Council (STFC) and the European Research Council (ERC). This work used the DiRAC@Durham facility managed by the
Institute for Computational Cosmology on behalf of the STFC DiRAC HPC Facility (www.dirac.ac.uk). The equipment was funded by BEIS capital funding via STFC capital grants ST/K00042X/1, ST/P002293/1, ST/R002371/1 and ST/S002502/1, Durham University and STFC operations
grant ST/R000832/1. DiRAC is part of the National e-Infrastructure. 
RG acknowledges financial support from the Spanish Ministry of Science and Innovation (MICINN) through the Spanish State Research Agency, under the Severo Ochoa Program 2020-2023 (CEX2019-000920-S). AF is supported by
the UK Research and Innovation (UKRI) Future Leaders Fellowships (grant numbers MR/V023381/1, MR/T042362/1).

\section*{Data Availability}

This study uses publicly available {\it Gaia} DR3 data.

\bibliography{references}

\appendix

\label{lastpage}

\end{document}